\documentclass[aps,prb,floatfix,superscriptaddress,twocolumn,footinbib]{revtex4-1}
\usepackage{subfigure}
\usepackage{amssymb}
\usepackage{amsmath}
\usepackage{amsfonts}
\usepackage{appendix}
\usepackage{bm}
\usepackage{graphicx}
\usepackage{epsfig}
\usepackage{epstopdf}
\usepackage{balance}
\usepackage[dvipsnames]{xcolor}
\usepackage{calc}
\usepackage{natbib}
\usepackage{color}
\usepackage[colorlinks,
            linkcolor=blue,
            anchorcolor=blue,
            citecolor=blue,
            urlcolor=blue]{hyperref}
\usepackage{lipsum}

\begin{document}

\title{Weak Localization and Antilocalization in
Twisted Bilayer Graphene}

\author{Hongyi Yan}
\affiliation{Center for Advanced Quantum studies, Department of Physics, Beijing Normal University, Beijing 100875, China}
\author{Haiwen Liu}
\email{haiwen.liu@bnu.edu.cn}
\affiliation{Center for Advanced Quantum studies, Department of Physics, Beijing Normal University, Beijing 100875, China}

\begin{abstract}
In this study, we investigate the weak localization (WL) and weak antilocalization (WAL) effects in twisted bilayer graphene positioned on a hexagonal boron nitride substrate. The bottom graphene layer aligns with the hexagonal boron nitride. The top layer of the system features a Dirac cone with a negligible gap, while the bottom layer possesses a  relatively large band gap. With a low concentration of impurities, the quantum correction to conductivity stems from the quantum interference between two time-reversed impurity scattering trajectories. We discover that inter-layer scattering significantly contributes to the conductivity correction when the Fermi surface areas of the two valleys at low energy are comparable. A double crossover from WL to WAL and back to WL occurs at a specific range of Fermi energy, which is particularly intriguing.\\
\end{abstract}

\maketitle

\section{Introduction}
Within the framework of classical mechanics, the Drude model is often used to describe electron motion and obtain the transport properties of materials. However, at low temperatures when the coherence length is sufficiently long and quantum effects cannot be ignored, the Drude model requires correction. Electrons in the material are scattered by impurities and form an infinite number of trajectories. The quantum interference between two closed time-reversed trajectories of electrons tends to increase the probability of electrons staying in their original place, which usually leads to a decrease in conductivity\cite{hikami1980spin,altshuler1980magnetoresistance}. This quantum interference can be suppressed by applying a magnetic field that breaks time-reversal symmetry and thereby breaks quantum coherence. The behavior of relativistic Dirac electrons in monolayer graphene is distinct from conventional electrons due to their linear energy band, large Fermi velocity, and chirality.\cite{novoselov2004electric}. Over the past decade, there has been extensive theoretical and experimental study of the quantum conductivity correction to the Drude model in monolayer \cite{ando1998berry, suzuura2002crossover, mccann2006weak, morozov2006strong, wu2007weak, kechedzhi2007weak, fal2007weak, yan2008weak, tikhonenko2008weak,tikhonenko2009transition,hilke2014weak} and bilayer graphene \cite{giovannetti2007substrate, kechedzhi2007weak, kechedzhi2007influence, horsell2008weak, liao2010gate}. Monolayer graphene typically exhibits weak antilocalization for long-ranged scattering impurities, while bilayer graphene shows weak localization. In addition to graphene systems, heterostructures \cite{amann2022counterintuitive, wang2015strong,wang2016origin, yang2016tunable, volkl2017magnetotransport, zihlmann2018large, wakamura2018strong}, semiconductors\cite{lu2013intervalley} and topological insulators \cite{imura2009weak,shan2012spin,lu2011weak,lu2011competition} are also used to study the phenomenon of weak localization and weak antilocalization.  Researchers have considered the influence of electron-electron interaction\cite{yan2008weak} spin-orbit coupling \cite{hikami1980spin, shan2012spin, lu2013intervalley, amann2022counterintuitive, wang2015strong, wang2016origin, yang2016tunable, volkl2017magnetotransport, zihlmann2018large, wakamura2018strong}, and magnetism\cite{shan2012spin,lu2011competition} and found that all of these factors have a significant and complex impact on transport properties.\\

In recent years, the discovery of the correlated insulating state and superconducting state in twisted bilayer graphene (TBG) near the magic angle \cite{cao2018correlated, cao2018unconventional} has made the twisted moir\'e system a hot topic of research\cite{lu2019superconductors,yankowitz2019tuning}. TBG consists of two graphene layers rotated at a specific angle $\theta$ to each other. which duplicates the original monolayer Brillouin zone's Dirac point into two equal Dirac points with the same chirality. \cite{bistritzer2011moire,dos2007graphene,moon2013optical,koshino2018maximally}. The interlayer hopping renormalizes the Fermi velocity at low energy, decreasing as the twist angle decreases. The key feature of TBG is that when the twist angle decreases to the so-called "magic angle", the bands near the charge neutral point are compressed into flat bands.  The kinetic energy of electrons on the bands dramatically reduces and becomes comparable to the strength of the Coulomb interaction, resulting in significant electron correlation that must be considered.  Currently, much attention has been focused on the study of strong correlation properties at the flat band, while weak (anti-)localization of TBG has not been extensively studied, with only a few experiments conducted on this subject \cite{meng2012enhanced,kashiwagi2022weak}. It remains unclear whether quantum correction of TBG or some artificial structure based on graphene has exotic characteristics compared to monolayer and bilayer graphene.

\begin{figure}
    \centering
    \includegraphics[width=8cm]{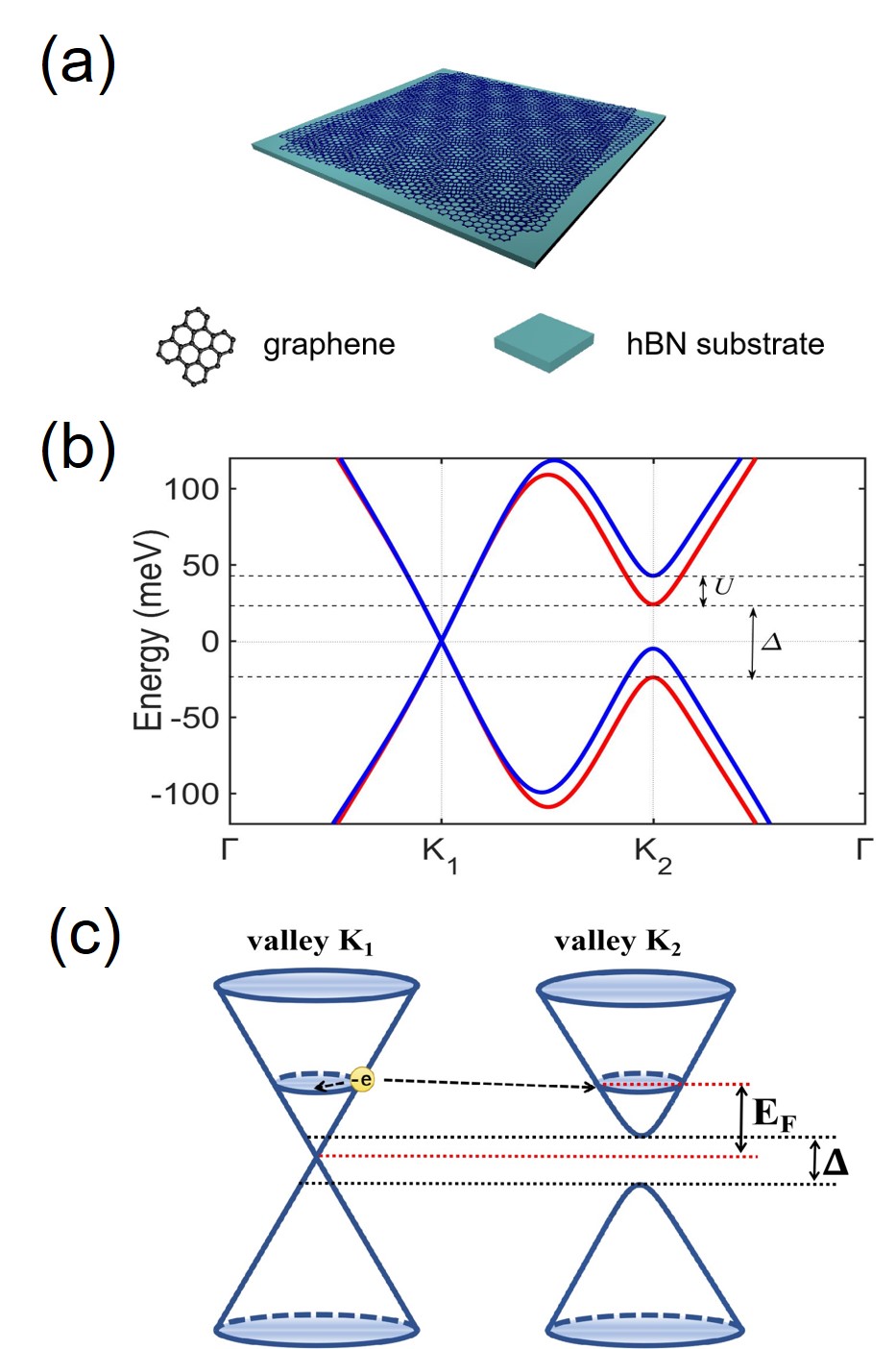}
    \caption{(a) Schematic diagram of the twisted bilayer graphene placed on hBN substrate system. (b) Blue (red) line is the band structure of the twisted bilayer graphene with (without) inter-layer bias, with the twist angle $\theta = 3^{\circ}$ . The band gap of the bottom layer is denoted as $\it \Delta$, and the bias-voltage induced bottom layer energy shift is represented by $\it U$. The two bands are put together in order to illustrate the bias voltage that causes the bottom layer band to shift. The Fermi surfaces of the system are nearly circular at the Fermi energy $E_F$ = 42 meV. (c) Low-energy band structure extracted from (b). Valley $\textbf{K}_{1}$ features a gapless and massless Dirac cone while valley $\textbf{K}_{2}$ is a massive energy band with a gap. The Fermi energy $E_{F}$ value is relative to the charge neutral point.}
    \label{figure 1}
\end{figure}

In this study, we present a theoretical investigation of the quantum conductivity correction in twisted bilayer graphene placed on a hexagonal boron nitride (hBN) substrate. The bottom layer of twisted bilayer graphene aligns with hBN. We consider a low concentration of long-range non-magnetic impurities as the main cause of electron scattering at low temperature and neglect electron-phonon and electron-electron scattering effects in the system. We particularly focus on the conductivity correction induced by inter-valley scattering between two valleys with different Fermi surface sizes (valley in this context refers to a mini-valley as described in previous work\cite{phinney2021strong}), which has not been studied previously. The paper is structured as follows. In Sec. \ref{II}, we introduce an effective Hamiltonian model for the system and the type of impurities considered. We then calculate the scattering time and velocity correction for intra- and inter-valley scattering. In Sec. \ref{III}, we carefully calculate the Cooperons for the two kinds of scattering. In Sec. \ref{IV}, we present our calculations for the quantum conductivity correction and magneto-conductivity. In Sec. \ref{V}, we summarize our findings and provide a discussion of our study. Finally in Sec. \ref{VI} we give our acknowledgement.

\section{Model} \label{II}
\subsection{Effective Hamiltonian}

The system under our study is twisted bilayer graphene placed in alignment on a hexagonal boron nitride substrate, as shown in Fig. \ref{figure 1}(a). Near the original $\textbf{K}$ point are two equal Dirac points derived from the top and bottom layers\cite{meyer2007structure}, labeled as $\textbf{K}_{1}$ and $\textbf{K}_{2}$, which we refer to as ``valleys'' instead of ``mini-valleys'' in this article. Since impurity scattering in the system is long-ranged, $\textbf{K}-\textbf{K}^{\prime}$ scattering is suppressed, we only need to consider scattering within one valley. The interaction between the hBN substrate and the bottom layer of graphene breaks the sublattice symmetry of the latter and results in a band gap opening whose magnitude depends on the on-site energy difference of the A and B sublattices of the bottom layer \cite{giovannetti2007substrate,
skomski2014sublattice}. We employ the modified Bistritzer-MacDonald continuum model \cite{bistritzer2011moire}, in which we set the on-site energy of the sublattices A and B in the bottom layer to be different to induce a band gap whose magnitude is $\Delta$. The band in the top layer can be considered as gapless and massless \cite{bistritzer2011moire} if we neglect the lattice relaxation effect. We also neglect  trigonal warping effect in the system since the Fermi surface is nearly circular in the energy regime of our consideration. These two effects do not influence our qualitative results. By tuning the bias voltage, one can control the relative energy shift $U$ between the two Dirac cones. The energy band of the system is depicted in Fig. \ref{figure 1}(b), and the simplified bands at low energy is extracted in Fig. \ref{figure 1}(c). The Van Hove singularity is located at the midpoint of the valleys $\textbf{K}_{1}$ and $\textbf{K}_{2}$. Its energy is positively correlated with the twist angle\cite{moon2013optical}. A larger twist angle corresponds to a higher Van Hove energy. When the Fermi energy approaches the Van Hove energy, the shape of the cone deforms such that we cannot use a simple Dirac equation to fit the band. Therefore, to obtain a relatively good cone shape,the twist angle should not be too small, and the Fermi energy should be far away from the Van Hove energy. Under these conditions, we can describe the low-energy band structure of the system using a simple effective Hamiltonian in the continuum limit, which is given by:

\begin{equation}\label{eq1}
H_{e f f}=\hat{\tau}_{0} \otimes \hbar v_F \textbf{k} \cdot \hat{\bm{\sigma}}+ \frac{\left(\hat{\tau}_{0}-\hat{\tau}_{z}\right)}{2} \otimes (\frac{\it \Delta}{2}\hat{\sigma}_{z}+U \hat{\sigma}_{0} ),
\end{equation}
$v_F$ is renormalized Fermi velocity, $ \hat{\bm{\sigma}}$, $\hat{\sigma}_{0}$ and $\hat{\sigma}_{z}$ are Pauli operators act in sublattice (A and B) space and $\hat{\tau}_{0}$ and $\hat{\tau}_{z}$ act in valley ($\textbf{K}_{1}$ and $\textbf{K}_{2}$) space. We then obtain the conduction band energy $E(\textbf{k}_{1})$ ($E(\textbf{k}_{2})$) and the corresponding eigenstates $\psi_1$ ($\psi_2$) near the valley $\textbf{K}_{1}$ ($\textbf{K}_{2}$). The wave function of the propagating electron is $\psi_{j}=\left\langle r \mid k_{j}\right\rangle=\frac{1}{\sqrt{S}}\left(a_{j}, b_{j} e^{i \phi_{\textbf{k}}}\right) e^{i \textbf{k}_{j} \cdot r}$. Here, $\phi_\textbf{k}$ is the phase angle of the wave vector, $j$ is the valley index. For two valleys $a_1=1/2$, $b_1=1/2$, $a_2=\sqrt{(E_F-U+\it \Delta/2)/2(E_F-U)}$, $b_2=\sqrt{(E_F-U-\it \Delta/2)/2(E_F-U)}$.

\subsection{Impurity type and scattering time}
We consider the case of weak non-magnetic impurity scattering, where perturbation treatment is applicable and electron spin does not flip. We investigate the transport properties of the system using standard weak (anti-)localization calculations. When the twist angle is small, the two valleys are in close proximity, and impurities can scatter electrons within the same valley or between the two valleys. Intra-valley scattering potential is given by the following expression:
\begin{equation}\label{eq2}
    U_{0}(\textbf{r})=\sum_{i} u_{0}^{i} \tau_{0} \otimes \sigma_{0} F\left(\textbf{r}-\textbf{R}_{i}\right).
\end{equation}
$F(\textbf{r}-\textbf{R}_{i})$ is the distribution function for long-ranged impurity scattering and can be simplified to the form of the delta function $\delta(\textbf{r}-\textbf{R}_{i})$\cite{lu2013intervalley}. Although this may seem contradictory at first sight because the delta function is short-ranged, its feasibility has been verified. Under the impurity averaging, $U_{0}(\textbf{r})$ satisfies $\left\langle U_{0}(\textbf{r})\right\rangle_{i m p}=0$, $\left \langle U_0(\textbf{r})U_0(\textbf{r}^{\prime}) \right \rangle _{imp} \sim \delta (\textbf{r}-\textbf{r}^{\prime})$. Intra-valley scattering amplitude is given by:

\begin{equation}\label{eq3}
\left\langle j ,\mathbf{k}\left|U_0\right| j , \mathbf{k}^{\prime}\right\rangle_{intra }=\frac{1}{S} \sum_i u_0^i e^{i\left(\mathbf{k}^{\prime}-\mathbf{k}\right) \cdot \mathbf{R}_i}\left(a^2_{j}+b^2_{j} e^{i\left(\phi_{\mathbf{k}^{\prime}}-\phi_{\mathbf{k}}\right)}\right).
\end{equation}
Intra-valley scattering time can be evaluated with the first Born approximation:

\begin{equation}\label{eq4}
\frac{1}{\tau}=\frac{2 \pi}{\hbar} \sum_{\mathbf{k}^{\prime}}\left\langle U_{\mathbf{k}, \mathbf{k}^{\prime}} U_{\mathbf{k}^{\prime}, \mathbf{k}}\right\rangle_{i m p} \delta\left(E_F-\varepsilon_{\mathbf{k}^{\prime}}\right),
\end{equation}
substitute Eq. (\ref{eq3}) into Eq. (\ref{eq4}) one can get:

\begin{equation}\label{eq5}
    \frac{1}{\tau_{jj}}=\frac{2 \pi N_{j,F}}{\hbar} n_{0} u_{0}^{2}\left(a_{j}^{4}+b_{j}^{4}\right),
\end{equation}
with $n_0u_0^2=\sum_{i}^{}(u_0^i)^2/S $. $n_0$ represents the concentration of non-magnetic impurities, $u_0$ is the average scattering potential, and $S$ is the area of the system. $N_{j,F}$ is the density of states (DOS) of valley $j$ at the Fermi energy per unit area of the system. Recent experimental studies have confirmed the existence of electron scattering between valleys in twisted bilayer graphene and twisted double bilayer graphene (TDBG)\cite{tomic2022scattering, phinney2021strong,kashiwagi2022weak}. Kashiwagi discovered that the ratio between intra-valley and inter-valley scattering time is proportional to $1/(\it \Delta K_\theta)^2$ in twisted double bilayer graphene\cite{kashiwagi2022weak}, where $\it \Delta K_\theta$ = $\left| \textbf{K}_{2}-\textbf{K}_{1}\right|$. We speculate same result can be found in TBG. Under this condition, inter-valley scattering amplitude is inversely proportional to $\it \Delta K_\theta$. We assume that inter-valley scattering potential has a similar form as in Eq.(\ref{eq2}),

\begin{equation}\label{eq6}
    U^{\prime}_{0}(\textbf{r})=\sum_{i} (u^{\prime})_{0}^{i} \tau_{x} \otimes \sigma_{0} F\left(\textbf{r}-\textbf{R}_{i}\right).
\end{equation}
where $u^{\prime}$ is inter-valley scattering strength which is usually smaller than intra-valley. Phinney found that when the twist angle $\theta = 1.65^{\circ}$, the ratio between intra-valley and inter-valley scattering time approximates 3/4\cite{phinney2021strong}. Therefore, We use the factor $\beta / \it \Delta K_\theta$ to express the ratio $(u^{\prime})_{0}^{i} /u_{0}^{i}$, so that the amplitude of the inter-valley scattering is inversely proportional to $\it \Delta K_\theta$,

\begin{equation}\label{eq7}
\begin{aligned}
\left\langle 1,\textbf{k}\left|U^{\prime}_{0}\right| 2,\textbf{k}^{\prime}\right\rangle_{\text {inter }}=&\frac{\beta}{\it \Delta K_{\theta}} \frac{1}{S} \sum_{i} u_{0}^{i} e^{i(\textbf{K}_2-\textbf{K}_1) \cdot \textbf{R}_i} e^{i\left(\textbf{k}^{\prime}-\textbf{k}\right) \cdot \textbf{R}_{i}} \\
&\times\left(a_{1}a_{2}+b_{1}b_{2} e^{i\left(\phi_{\textbf{k}^{\prime}}-\phi_{\textbf{k}}\right)}\right).
\end{aligned}
\end{equation}
Using Eq. (\ref{eq4}) inter-valley scattering time can be calculated,

\begin{equation}\label{eq8}
    \frac{1}{\tau_{j\bar{j}}}=\frac{2 \pi N_{\bar{j},F}}{\hbar} \frac{\beta^2}{\left(\it \Delta K_{\theta}\right)^{2}} n_{0} u_{0}^{2}\left(a_{j}^{2} a_{\bar{j}}^{2}+b_{j}^{2} b_{\bar{j}}^{2}\right).
\end{equation}
here $j$ = 1, 2, $\bar{j}$ is another valley relative to $j$. $\tau_{j\bar{j}}$ denotes the scattering time of electron scattered from valley $j$ to valley $\bar{j}$. The scattering time is inversely proportional to the DOS of the destination. The total scattering time of the two valleys are:

\begin{equation}\label{eq9}
\begin{aligned}
&\frac{1}{\tau_{1}}=\frac{1}{\tau_{11}}+\frac{1}{\tau_{12}}, \\
&\frac{1}{\tau_{2}}=\frac{1}{\tau_{22}}+\frac{1}{\tau_{21}}.
\end{aligned}
\end{equation}


\subsection{Velocity correction}
Firstly, we consider the Diffuson approximation, which is a vertex correction that describes the multiple scattering of electrons on the impurity potential \cite{akkermans2007mesoscopic,datta1997electronic}. The Diffuson correction corresponds to a series of ladder diagrams and contributes to the velocity correction through an iterative equation \cite{altshuler1980magnetoresistance, shan2012spin}:

\begin{equation}\label{eq10}
    \widetilde{v}_{j,\textbf{k}}^{x}=v_{j,\textbf{k}}^{x}+\sum_{j^{\prime}, \textbf{k}^{\prime}} G_{j^{\prime}, \textbf{k}^{\prime}}^{R} G_{j^{\prime}, \textbf{k}^{\prime}}^{A}\left\langle U_{\textbf{k},\textbf{k}^{\prime}} U_{\textbf{k}^{\prime},\textbf{k}} \right\rangle_{i m p} \widetilde{v}_{j^{\prime}, \textbf{k}^{\prime}}^{x},
\end{equation}

\begin{equation}\label{eq11}
G_{j,\textbf{k}}^{R / A}=\frac{1}{\omega-\varepsilon_{j,\textbf{k}} \pm i \hbar / 2 \tau_{j}}.
\end{equation}
${v}_{j, \textbf{k}}^{x}$, $\widetilde{v}_{j, \textbf{k}}^{x}$ respectively denote bare velocity and corrected velocity of valley $j$ in the $x$ direction, $v^{x}=vcos\phi_{\textbf{k}}$. Eq. (\ref{eq11}) is the expression for the retarded and advanced Green's functions, corresponding to the plus and minus signs in the denominator on the right-hand side of the equation. Eq. (\ref{eq10}) can be further expanded and written as a set of equations:

\begin{equation}\label{eq12}
    \begin{aligned}
v_{1,\textbf{k}}^{x}=&\left(1-\frac{\tau_{1}}{\tau_{11}} \frac{a_{1}^{2} b_{1}^{2}}{a_{1}^{4}+b_{1}^{4}}\right)\widetilde{v}_{1,\textbf{k}}^{x} -\frac{\tau_{2}}{\tau_{12}} \frac{a_{1} a_{2} b_{1} b_{2}}{a_{1}^{2} a_{2}^{2}+b_{1}^{2} b_{2}^{2}} \widetilde{v}_{2,\textbf{k}_{//}}^{x}, \\
v_{2,\textbf{k}}^{x}=&\left(1-\frac{\tau_{2}}{\tau_{22}} \frac{a_{2}^{2} b_{2}^{2}}{a_{2}^{4}+b_{2}^{4}}\right)\widetilde{v}_{2,\textbf{k}}^{x} -\frac{\tau_{1}}{\tau_{21}} \frac{a_{1} a_{2} b_{1} b_{2}}{a_{1}^{2} a_{2}^{2}+b_{1}^{2} b_{2}^{2}} \widetilde{v}_{1,\textbf{k}_{//}}^{x}. \\
\end{aligned}
\end{equation}
$\textbf{k}_{//}$ denotes the wave vector in the valley $\textbf{K}_{2}$ ($\textbf{K}_{1}$) on the Fermi surface that is parallel to the wave vector $\textbf{k}$ in the valley $\textbf{K}_{1}$ ($\textbf{K}_{2}$). Note that the two bands are isotropic, the Fermi velocities on their own Fermi surface are of equal magnitudes and independent of directions. By numerically solving Eq. (\ref{eq12}), we can obtain the velocity correction factor $\eta_{j}$:

\begin{equation}\label{eq13}
\begin{aligned}
\eta_{1}=&\widetilde{v}_{1,\textbf{k}}^{x}/ v_{1,\textbf{k}}^{x}, \\
\eta_{2}=&\widetilde{v}_{2,\textbf{k}}^{x}/ v_{2,\textbf{k}}^{x}.
\end{aligned}
\end{equation}


\section{Cooperon correction}\label{III}

\subsection{Cooperon}
\begin{figure*}
    \centering
    \includegraphics[width=14cm,height=15cm]{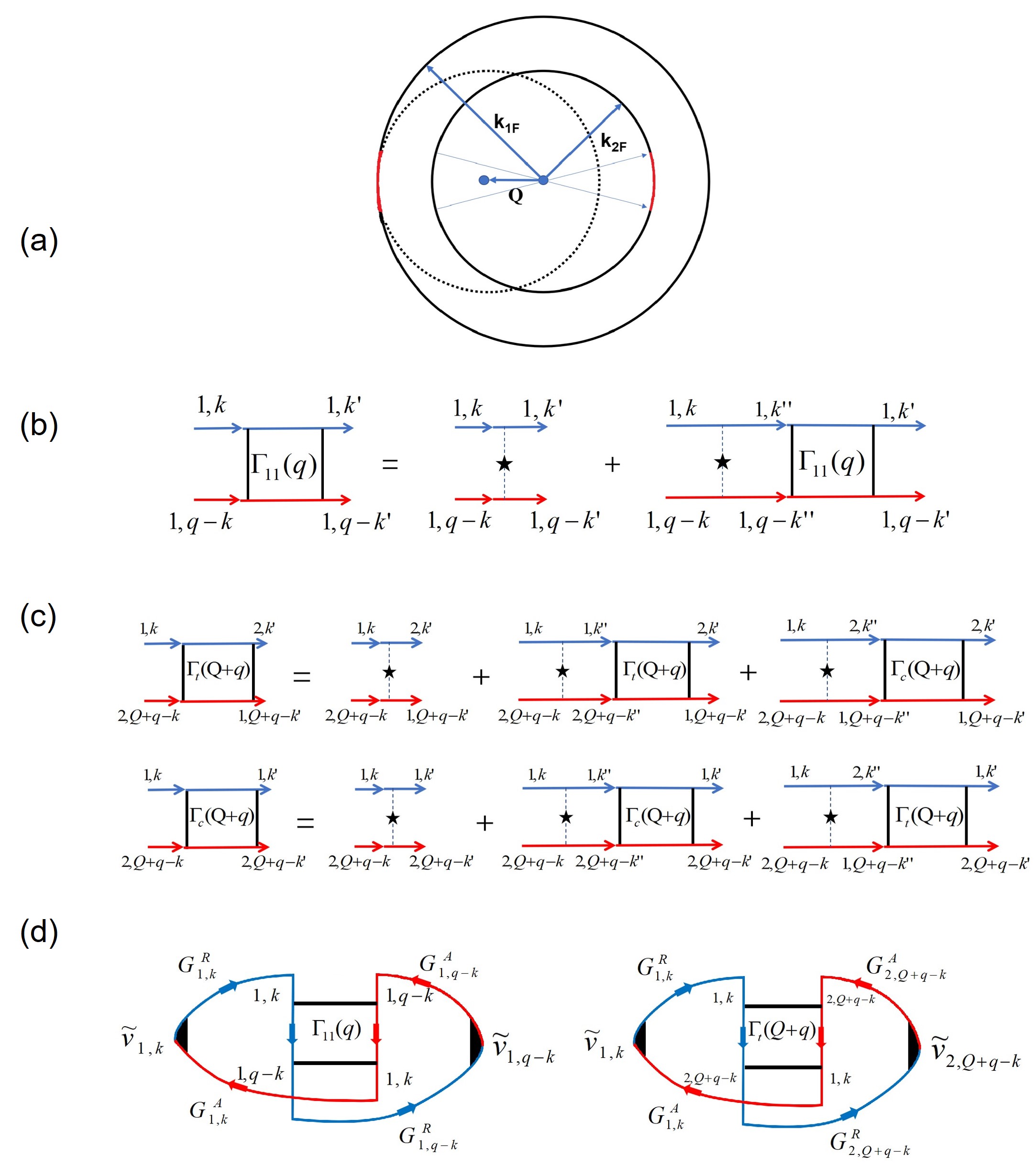}
    \caption{(a) The Fermi surface of the system at a specific energy. To aid visualization, we shift the Fermi circles of the two valleys to the same center. The larger (smaller) circle corresponds to the Fermi surface of valley $\textbf{K}_{1}$ ($\textbf{K}_{2}$). $\textbf{k}_{1F}$ and $\textbf{k}_{2F}$ denote the Fermi wave vectors on the Fermi circle of the two valleys. The nesting wave vector $\textbf{Q}$ has a modulus of $\left|\textbf{Q}\right|=\left|\textbf{k}_{1F}\right|-\left|\textbf{k}_{2F}\right|$. The red arcs on the two Fermi circles indicate the part of wave vector $\textbf{k}$ on the larger circle and $\textbf{k}^{\prime}$ on the smaller circle that satisfies $\textbf{k} +\textbf{k}^{\prime} = \textbf{Q} + \textbf{q}$ and contribute to the conductivity correction. (b) The Bethe-Salpeter equation that the intra-valley scattering Cooperon $\Gamma_{11}$ satisfies. The blue (red) line represents the retarded (advanced) Green function. The pentacle denotes an impurity, and impurity scattering is shown by dotted lines. (c) The Bethe-Salpeter equations that the ``trans'' and ``cis'' Cooperons satisfy. (d) The Feynman diagram depicts the intra- and inter- valley scattering contribution of conductivity correction.}
    \label{figure 2}
\end{figure*}

Next, we consider the Cooperon correction that corresponds to the so-called ``maximum-crossed diagrams'' \cite{akkermans2007mesoscopic,datta1997electronic}, which describes the quantum interference between two time-reversal electron wave trajectories of oppositely ordered collision sequences. Although for intra-valley scattering, the wave vectors $\textbf{k}$ and $-\textbf{k}$ are not strictly time-reversal because they are relative to the Dirac point $\textbf{K}$ rather than $\Gamma$ in the Brillouin zone of monolayer graphene, we still consider them to satisfy generalized time reversion. The Cooperon directly contributes to the conductivity correction as follows:

\begin{equation}\label{eq14}
\begin{aligned}
    \it \Delta \sigma=&\frac{e^{2} \hbar}{2 \pi S} \sum_{j,j^{\prime},\textbf{k}} \sum_{\textbf{q}} \Gamma_{j j^{\prime}}(\textbf{Q}+\textbf{q}) \widetilde{v}_{j,\textbf{k}}^{x} \widetilde{v}_{j^{\prime},\textbf{Q}+\textbf{q}-\textbf{k}}^{x}  \\
&\times G_{j,\textbf{k}}^{R}  G_{j,\textbf{k}}^{A} G_{j^{\prime},\textbf{Q}+\textbf{q}-\textbf{k}}^{R} G_{j^{\prime},\textbf{Q}+\textbf{q}-\textbf{k}}^{A}.  \\
\end{aligned}
\end{equation}
$\Gamma_{jj^{\prime}}(\textbf{Q}+\textbf{q})$ is the Cooperon corresponding to the scattering from valley $j$, vector $\textbf{k}$ to valley $j^{\prime}$, vector $\textbf{k}^{\prime}$. $j$, $j^{\prime}$ = 1, 2 and $\textbf{k}+\textbf{k}^{\prime}=\textbf{Q}+\textbf{q}$. Here, $\textbf{Q}$ is defined as a ``nesting vector" for inter-valley scattering. Its modulus equals the radius difference between the two Fermi circles, ${\left|\textbf{Q}\right|=\left|\textbf{k}_{1F}\right|-\left|\textbf{k}_{2 F}\right|}$, as shown in the Fig. \ref{figure 2}(a). For intra-valley scattering, the nesting vector is $\textbf{Q} =0$. To obtain the Cooperon $\Gamma(\textbf{Q}+\textbf{q}) $, we need to evaluate the bare vertex function $\gamma$. The bare vertex function can be calculated by:

\begin{equation}\label{eq15}
\gamma_{j,\textbf{k}; j^{\prime}, \textbf{k}^{\prime}} \equiv\left\langle U_{j, \textbf{k} ; j^{\prime}, \textbf{k}^{\prime}} U_{j^{\prime}, \textbf{Q}+\textbf{q}-\textbf{k} ; j, \textbf{Q}+\textbf{q}-\textbf{k}^{\prime}}\right\rangle_{i m p}. \\
\end{equation}
In the following, we will provide a detailed expression of the bare vertex function $\gamma$ for intra-valley and inter-valley scattering, and then calculate the corresponding Cooperons.\\

To begin with, we calculate the Cooperon for intra-valley scattering ($j=j^\prime$). By combining Eqs. (\ref{eq3}) and (\ref{eq15}), and setting the nesting wave vector $\textbf{Q}=0$, we can calculate the bare vertex function for intra-valley scattering as follows:

\begin{equation}\label{eq16}
    \begin{aligned}
\begin{aligned}
\gamma_{j,\textbf{k};j,\textbf{k}^{\prime}} & \equiv\left\langle U_{j, \mathbf{k} ; j, \mathbf{k}^{\prime}} U_{j, \mathbf{q}-\mathbf{k} ; j, \mathbf{q}-\mathbf{k}^{\prime}}\right\rangle_{i m p} \\
&=\frac{n_0 u_0^2}{S} \{ a_j^4+2 a_j^2 b_j^2 e^{i\left(\phi_{\mathbf{k}^{\prime}}-\phi_{\mathbf{k}}\right)}+b_j^4 e^{2 i\left(\phi_{\mathbf{k}^{\prime}}-\phi_{\mathbf{k}}\right)} \}
\end{aligned}
\end{aligned}
\end{equation}
In Eq. (\ref{eq16}), the bare vertex function has three channels $L$ = 0, 1, 2, corresponding to the three terms separately. Here, $L$ is the coefficient of $\phi_{\textbf{k}^{\prime}}-\phi_{\textbf{k}}$ in the exponent. To calculate the quantum correction to the conductivity, we set $\phi_{\textbf{k}^{\prime}}-\phi_{\textbf{k}}=\pi$ and $\textbf{k}+\textbf{k}^{\prime} = \textbf{q}$. The Cooperon is related to the bare vertex function through the Bethe-Salpeter equation \cite{akkermans2007mesoscopic}:

\begin{equation}\label{eq17}
\Gamma_{j,\textbf{k};j,\textbf{k}^{\prime}}=\gamma_{j,\textbf{k};j,\textbf{k}^{\prime}}+\sum_{\textbf{k}^{\prime \prime}} \gamma_{j,\textbf{k};j,\textbf{k}^{\prime \prime}} G_{j,\textbf{k}^{\prime \prime}}^{R} G_{j,\textbf{q}-\textbf{k}^{\prime \prime}}^{A} \Gamma_{j,\textbf{k}^{\prime \prime};j,\textbf{k}^{\prime}}.
\end{equation}
The Feynman diagram representation of this equation is depicted in Fig. \ref{figure 2}(b). Similar to the bare vertex function $\gamma$, the Cooperon $\Gamma$ also has three channels denoted by $L$ = 0, 1, 2:

\begin{equation}\label{eq18}
\begin{aligned}
\Gamma_{j,\textbf{k};j,\textbf{k}^{\prime}}&= \Gamma_{jj}(\textbf{q}) \\
&=\Gamma_{j}^{(0)}+\Gamma_{j j}^{(1)} e^{i\left(\phi_{\textbf{k}^{\prime}}-\phi_{\textbf{k}}\right)}+\Gamma_{j j}^{(2)} e^{2 i\left(\phi_{\textbf{k}^{\prime}}-\phi_{\textbf{k}}\right)}.
\end{aligned}
\end{equation}
Intra-valley scattering Cooperon is written in a simple form as $\Gamma_{jj}(\textbf{q})$. We assume that the three channels obey the Bethe-Salpeter equation separately and do not couple with each other \cite{imura2009weak}. In Eq.(\ref{eq17}), the product of $\gamma_{j,\textbf{k} ;j,\textbf{k}^{\prime \prime}} \Gamma_{j,\textbf{k}^{\prime \prime};j,\textbf{k}^{\prime } }$ does not contain the parameter $\phi_{\textbf{k}^{\prime \prime}}$ as it is canceled out in the exponent. Therefore, this term can be taken out of the summation over $\textbf{k}^{\prime \prime}$. We can then define the summation as $\Pi$, transform the summation of $\textbf{k}^{\prime\prime}$ into an integral, and evaluate it to obtain:

\begin{equation}\label{eq19}
\Pi_{j j}=\sum_{\textbf{k}^{\prime \prime}} G_{j,\textbf{k}^{\prime \prime}}^{R} G_{j,\textbf{q}-\textbf{k}^{\prime \prime}}^{A}=\frac{2 \pi N_{j,F}}{\hbar} \tau_{j}\left(1-D_{j} q^{2} \tau_{j}\right),
\end{equation}
here $\tau_j$ is the total scattering time of the valley $j$ in the Eq. (\ref{eq9}), $D_{j}$ is diffusion constant with $D_{j}= {v}_{j}\tau_{j}^{2}/2$. Eq. (\ref{eq17}) can be transformed into a concise form:

\begin{equation}\label{eq20}
    \Gamma_{j j}=\gamma_{j j}+\Pi_{j j} \gamma_{j j} \Gamma_{j j}.
\end{equation}
$\gamma_{jj}$ is a concise form of the bare vertex function. Solve the Eq. (\ref{eq20}), the Cooperon $\Gamma_{jj}$ can be obtained:

\begin{equation}\label{eq21}
    \Gamma_{j j}=\frac{\gamma_{j j}}{1-\prod_{j j} \gamma_{j j}}.
\end{equation}
Combining Eqs. (\ref{eq18})-(\ref{eq23}), we can obtain an explicit expression of the Cooperon:

\begin{equation}\label{eq22}
    \Gamma_{j j}^{(L)}=\frac{1}{Z_{j} D_{j} \tau_{j}^{2}} \frac{1}{l_{j,L}^{-2}+q^{2}},
\end{equation}

\begin{equation}\label{eq23}
    l_{j,L}^{-2}=\frac{1-Z_{j} \tau_{j} \gamma_{j j}^{(L)}}{Z_{j} \tau_{j}^{2} D_{j} \gamma_{j j}^{(L)}}.
\end{equation}
In Eqs. (\ref{eq22}) and (\ref{eq23}), $Z_{j}=2\pi N_{j,F}/\hbar $, where $N_{j,F}$ is the density of states of the $j$ valley at the Fermi level, and $l_{L}$ is the scattering length of channel $L$. Since there are always three channels $L$ = 0, 1, 2 for Cooperon, we omit the channel index $L$ for brevity in the following. It should be noted that all six channels from the two valleys have a finite scattering length. The gap in the Cooperon of channel $L$ = 1 of the valley $\textbf{K}_1$ results from inter-valley scattering, which changes the scattering time of the valley according to Eq. (\ref{eq9}). This differs from the situation in monolayer graphene, where the $L$ = 1 channel corresponds to a Goldstone mode when only intra-valley scattering occurs \cite{imura2009weak}.

When considering the inter-layer scattering Cooperon, the situation becomes more complicated because the Fermi surface of the two valleys are not equal in size. For any given wave vector $\textbf{k}$ in valley $\textbf{K}_1$, we cannot find $-\textbf{k}$ in valley $\textbf{K}_2$. Thus, we introduce a non-zero nesting vector $\textbf{Q}$. The primary contribution to the conductivity correction comes from the portion of wave vectors $\textbf{k}$ on the Fermi surface of one valley that scatter to the wave vector $\textbf{Q}+\textbf{q}-\textbf{k}$ of the other valley. This proportion is represented by the red arcs in Fig. \ref{figure 2}(a). The bare vertex function for inter-valley scattering is expressed as:

\begin{equation}\label{eq24}
\begin{aligned}
\gamma_{12,21} \equiv&\left\langle U_{1, \textbf{k}; 2, \textbf{k}^{\prime}} U_{2, \textbf{Q}+\textbf{q}-\textbf{k} ; 1, \textbf{Q}+\textbf{q}-\textbf{k}^{\prime}}\right\rangle_{i m p}\\
=&\frac{\beta^2}{(\it \Delta K_{\theta})^{2}} \frac{n_{0} u_{0}^{2}}{S}
\{ a_{1}^{2} a_{2}^{2} +2a_{1} a_{2} b_{1} b_{2} e^{i\left(\phi_{\textbf{k}^{\prime}}-\phi_{\textbf{k}}\right)} \\
&+b_{1}^{2} b_{2}^{2} e^{2 i\left(\phi_{\textbf{k}^{\prime}}-\phi_{\textbf{k}}\right)} \}.
\end{aligned}
\end{equation}
The subscripts 12,21 of $\gamma$ represent the correlation between the scattering events of electrons from valley $\textbf{K}_{1}$ to valley $\textbf{K}_{2}$ and its time-reversed process. ``12'' denotes $1,\textbf{k}$ and $2,\textbf{Q}+\textbf{q}-\textbf{k}$ on the upper (blue) and lower (red) electron line respectively before scattering event happens as seen in Fig. \ref{figure 2}(c). To distinguish inter-valley scattering bare vertex function and Cooperon from those of intra-valley in Eqs. (\ref{eq16}) and (\ref{eq18}), we use four indices to represent them as shown in Eq. (\ref{eq24}). To calculate the Cooperon $\Gamma_{12,21}$, we need another Cooperon $\Gamma_{12,12}$ seen in the second line of Fig. \ref{figure 2}(c). The bare vertex function of $\gamma_{12,12}$ is defined as:

\begin{equation}\label{eq25}
\begin{aligned}
\gamma_{12,12} \equiv&\left\langle U_{1, \textbf{k}; 1, \textbf{k}^{\prime}} U_{2, \textbf{Q}+\textbf{q}-\textbf{k} ; 2, \textbf{Q}+\textbf{q}-\textbf{k}^{\prime}}\right\rangle_{i m p}\\
=& \frac{n_{0} u_{0}^{2}}{S} \{ a_{1}^{2} a_{2}^{2} +(a_{1}^{2} b_{2}^{2}+ a_{2}^{2} b_{1}^{2}) e^{i\left(\phi_{\textbf{k}^{\prime}}-\phi_{\textbf{k}}\right)} \\
&+b_{1}^{2} b_{2}^{2} e^{2 i\left(\phi_{\textbf{k}^{\prime}}-\phi_{\textbf{k}}\right)} \}.
\end{aligned}
\end{equation}
The form of $\gamma_{12,12}$ is similar to that of intra-valley scattering in Eq. (\ref{eq16}). However, it represents the correlation between intra-valley scattering events in valley $\textbf{K}_{1}$ and $\textbf{K}_{2}$, and its meaning is completely different from that of Eq. (\ref{eq16}). Additionally, the Cooperon of this type, $\Gamma_{12,12}$, does not contribute to the conductivity correction because the two wave vectors on the diagonal of the Cooperon, $1,\textbf{k}$ and $2,\textbf{Q}+\textbf{q}-\textbf{k}^{\prime}$, are not equal. As a result, the Bethe-Salpeter equation takes a coupled form:

\begin{equation}\label{eq26}
\begin{aligned}
\Gamma_{12, 12}=&\gamma_{12,12}+ \sum_{\textbf{k}^{\prime \prime}}^{\prime} \gamma_{12,12} G_{1, \textbf{k}^{\prime \prime}}^{R} G_{2, \textbf{Q}+\textbf{q}-\textbf{k}^{\prime \prime}}^{A} \Gamma_{12,12}\\
&+\sum_{\textbf{k}^{\prime \prime}}^{\prime} \gamma_{12,21} G_{2, \textbf{k}^{\prime \prime}}^{R} G_{1, \textbf{Q}+\textbf{q}-\textbf{k}^{\prime \prime}}^{A} \Gamma_{21,12}.  \\
\Gamma_{21,12}=&\gamma_{21,12}+ \sum_{\textbf{k}^{\prime \prime}}^{\prime} \gamma_{21,12} G_{1,\textbf{k}^{\prime \prime}}^{R} G_{2, \textbf{Q}+\textbf{q}-\textbf{k}^{\prime \prime}}^{A} \Gamma_{12,12} \\
&+\sum_{\textbf{k}^{\prime \prime}}^{\prime} \gamma_{21,21} G_{2, \textbf{k}^{\prime \prime}}^{R} G_{1, \textbf{Q}+\textbf{q}-\textbf{k}^{\prime \prime}}^{A} \Gamma_{21,12}.
\end{aligned}
\end{equation}
A prime on the summation sign in Eq. (\ref{eq26}) indicates that the summation over the wave vector $\textbf{k}^{\prime \prime}$ is not over whole points on the Fermi surface but over the points on the arc. The Feynman diagram representation of this set of equations is shown in Fig. \ref{figure 2}(c). The product of $\gamma\Gamma$ does not contain $\phi_{\textbf{k}^{\prime\prime}}$, so it can be factored out of the summation, leaving the summation similar to Eq. (\ref{eq19}), which we define as the summation $\Pi_{j\bar{j}}$:

\begin{equation}\label{eq27}
\Pi_{j \bar{j}}=\sum_{\textbf{k}^{\prime \prime}}^{\prime} G_{j,\textbf{k}^{\prime \prime}}^{R} G_{\bar{j},\textbf{Q}+\textbf{q}-\textbf{k}^{\prime \prime}}^{A},
\end{equation}

To calculate $\Pi_{j \bar{j}}$ exactly, two things need to be clarified. Firstly, only the $\textbf{k}^{\prime\prime}$ points near the tangent point and on the red arcs in Fig. \ref{figure 2}(a) are considered in the calculation. By introducing the nesting wave vector $\textbf{Q}$, we match as many wave vectors $\textbf{k}$ of one valley and $\textbf{k}^{\prime}$ of another valley as possible, such that they satisfy $\textbf{k}=\textbf{Q}+\textbf{q}-\textbf{k}^{\prime \prime}$ and all wave vectors on the two arcs correspond one to one. Due to the isotropy of the two bands, we suppose to approximate the partial summation by summing over all points on the Fermi surface multiplied by a hyperbolic tangent function $f = \tanh(\lambda k_{min}/Q)$, where $k_{min}$ is the modulus of the smaller Fermi wave vectors of the two valleys. Secondly, the summations $\Pi_{12}$ and $\Pi_{21}$ are equal. Since they correspond to the same wave vector $\textbf{k}^{\prime\prime}$ summation, and they are conjugate to each other. Furthermore, considering the fact that they are real numbers, we can assume that $\Pi_{12}$ and $\Pi_{21}$ are equal. Let $\Pi_{12} = \Pi_{21} = \Pi$. With the above analysis, the form of $\Pi$ can be approximated as:

\begin{equation}\label{eq28}
\Pi=tanh(\lambda \frac {k_{min}}{Q})\frac{2 \pi N}{\hbar} \tau \left(1-D q^{2} \tau \right).
\end{equation}
In Eq. (\ref{eq28}), we have already used the condition $\epsilon_{j,\textbf{k}}=\epsilon_{\bar{j},\textbf{Q}-\textbf{k}}$. Here, $N$ represents the average density of states of the two valleys at Fermi energy, given by $N=(N_{1,F}+N_{2,F})/2$. The scattering time $\tau$ satisfies $2/\tau=1/\tau_{1}+1/\tau_{2}$, where $\tau_1$ and $\tau_2$ are the scattering times in valleys $\textbf{K}_1$ and $\textbf{K}_2$ respectively. The diffuson constant $D$ is defined as $D=v^{2}\tau/2$, where the velocity $v$ satisfies $v^{2}= (N_{1,F}v_{1}^{2}+N_{2,F}v_{2}^{2})/(N_{1,F}+N_{2,F})$. This treatment mainly takes the average of the two valleys in order to obtain the equal $\Pi_{12}$ and $\Pi_{21}$. The Bethe-Salpeter equation in Eq. (\ref{eq26}) can be modified into a set of equations:

\begin{equation}\label{eq29}
    \left(\begin{array}{l}
\Gamma_{12,12} \\
\Gamma_{21,12}
\end{array}\right)=\left(\begin{array}{l}
\gamma_{12,12} \\
\gamma_{21,12}
\end{array}\right)+\left(\begin{array}{ll}
\gamma_{12,12}  & \gamma_{12,21}  \\
\gamma_{21,12} & \gamma_{21,21}
\end{array}\right)\Pi\left(\begin{array}{l}
\Gamma_{12,12} \\
\Gamma_{21,12}
\end{array}\right).
\end{equation}
Furthermore, the bare vertex functions satisfy the following relations: $\gamma_{12,12} = \gamma_{21,21} = \gamma_{c}$, $\gamma_{12,21} = \gamma_{21,12} = \gamma_{t}$, where $\gamma_{c}$ and $\gamma_{t}$ represent the ``cis" and ``trans" forms, respectively\cite{imura2009weak}. We can rewrite $\Gamma_{12,12}$ as $\Gamma_{c}$ and $\Gamma_{21,12}$ as $\Gamma_{t}$, and simplify Eq. (\ref{eq29}) as:

\begin{equation}\label{eq30}
    \left(\begin{array}{c}
\Gamma_{c} \\
\Gamma_{t}
\end{array}\right)=\left(\begin{array}{l}
\gamma_{c} \\
\gamma_{t}
\end{array}\right)+\left(\begin{array}{ll}
\gamma_{c} & \gamma_{t} \\
\gamma_{t} & \gamma_{c}
\end{array}\right)\Pi \left(\begin{array}{c}
\Gamma_{c} \\
\Gamma_{t}
\end{array}\right)
\end{equation}
Only $\Gamma_{t}$ contributes to the conductivity correction and it can be solved with the form:

\begin{equation}\label{eq31}
\begin{aligned}
\Gamma_{t}=\frac{\gamma_{t}}{1-2 \Pi\gamma_{c} +\Pi^{2}\left(\gamma_{c}^{2}-\gamma_{t}^{2}\right) }.
\end{aligned}
\end{equation}
 Different from the Cooperon solved in Eq. (\ref{eq22}), the Cooperons in Eq. (\ref{eq31}) cannot be directly transformed to a standard form of $1/(l^{-2}+q^{2})$. At this point, we can neglect the $q^4$ term in the denominator, which is a small quantity of higher order. As a result, the Cooperon $\Gamma_t$ can be expressed as:

\begin{equation}\label{eq32}
\begin{aligned}
& \Gamma_{t}=\alpha_{t} \frac{1}{ZD\tau^{2}} \frac{1}{l_{t}^{-2}+q^{2}}\\
& \alpha_{t}^{-1}=2\frac{\gamma_{c}}{\gamma_{t}}-2Z\tau \frac{\gamma_{c}^{2}-\gamma_{t}^{2}}{\gamma_{t}}\\
& l_{t}^{-2}=\frac{1}{D}(\frac{1}{Z \tau^{2} \gamma_{t}}-{2 \frac{\gamma_{c}}{\tau_{c} \gamma_{t}}}+Z \frac{\gamma_{c}^{2}-\gamma_{t}^{2}}{\gamma_{t}}).
\end{aligned}
\end{equation}
$l_{t}$ is the scattering length of inter-valley scattering Cooperon. $Z=2\pi N\cdot\tanh(\lambda k_{min}/Q)/\hbar$. The form of the Cooperon is analogous to that of intra-valley scattering, as shown in Eq. (\ref{eq22}).
Notice that each expression contains three channels.

\section{Conductivity correction versus temperature and magnetic field}\label{IV}

\begin{figure*}
    \centering
    \includegraphics[width=16cm,height=9cm]{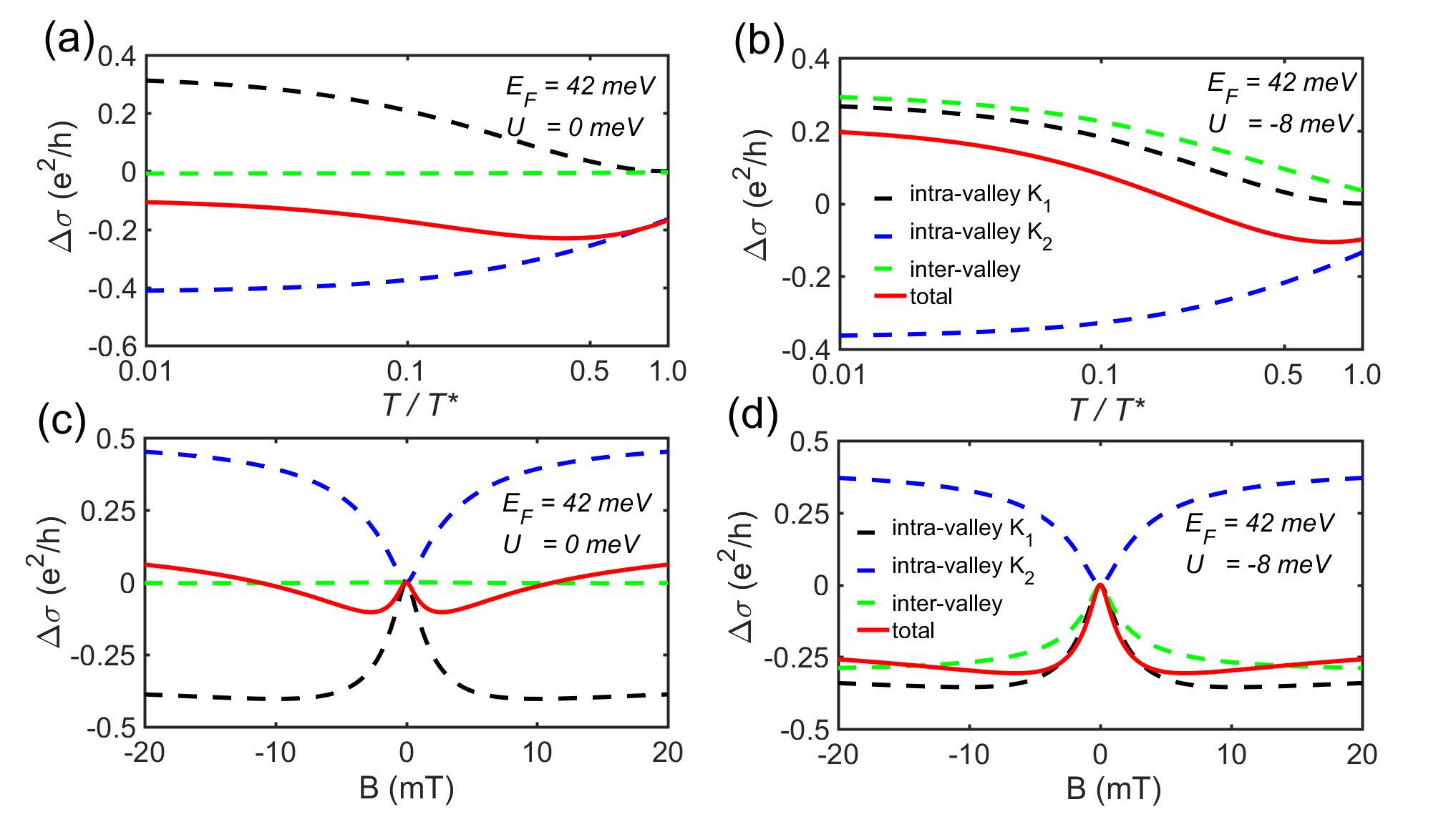}
    \caption{Conductivity correction and magnetoconductivity plots. (a) Conductivity correction of the system at the Fermi energy $E_{F}$ = 42 meV without bias. The gap $\it \Delta$ is set to be 53 meV. The red solid line denotes the total correction, while the black, blue, and green dotted lines represent the contribution of the corresponding parts. (b) Same parameter used as (a) except for an energy shift $U$ = - 8 meV. (c) Magneto-conductivity plot corresponds to (a). (d) Magneto-conductivity plot corresponds to (b). The mean free path $l_{1,e}$ is set to be 400 nm, and the temperature $T/T^{*}$ = 0.01. We observe a crossover from weak localization to weak anti-localization at low temperatures.}
    \label{figure 3}
\end{figure*}

The conductivity correction is directly connected with the Cooperon through:

\begin{equation}\label{eq33}
\begin{aligned}
\it \Delta\sigma_{intra}=& \frac{e^{2}\hbar}{2\pi }\sum_{j} \sum_{\textbf{q}} \Gamma_{jj}(\textbf{q}) \sum_{\textbf{k}} \widetilde{v}_{j, \textbf{k}}^{x} \widetilde{v}_{j, \textbf{q}-\textbf{k}}^{x} \\
& \times G_{j, \textbf{k}}^{R} G_{j, \textbf{k}}^{A} G_{j, \textbf{q}-\textbf{k}}^{R} G_{j, \textbf{q}-\textbf{k}}^{A},
\end{aligned}
\end{equation}

\begin{equation}\label{eq34}
\begin{aligned}
\it \Delta\sigma_{inter}=& \frac{e^{2}\hbar}{2\pi } \sum_{j} \sum_{\textbf{q}} \Gamma_{j\bar{j}}(\textbf{Q+q}) \sum_{\textbf{k}} ^{\prime} \widetilde{v}_{j, \textbf{k}}^{x} \widetilde{v}_{\bar{j}, \textbf{Q+q}-\textbf{k}}^{x} \\
& \times G_{j, \textbf{k}}^{R} G_{j, \textbf{k}}^{A} G_{\bar{j}, \textbf{Q+q}-\textbf{k}}^{R} G_{\bar{j}, \textbf{Q+q}-\textbf{k}}^{A}.
\end{aligned}
\end{equation}
Eqs. (\ref{eq33}) and (\ref{eq34}) represent the conductivity correction contributions from intra- and inter- valley scattering. To begin with, we calculate the summation of $\textbf{k}$:

\begin{equation}\label{eq35}
\sum_{\textbf{k}} \widetilde{v}_{j, \textbf{k}}^{x} \widetilde{v}_{j, \textbf{q}-\textbf{k}}^{x} G_{j, \textbf{k}}^{R} G_{j, \textbf{k}}^{A} G_{j, \textbf{q}-\textbf{k}}^{R} G_{j, \textbf{q}-\textbf{k}}^{A}
=- \frac{2 \pi \eta_{j}^{2}v_{j}^{2}N_{j,F}\tau_{j}^{3}}{\hbar^{3}},
\end{equation}

\begin{equation}\label{eq36}
\begin{aligned}
&\sum_{\textbf{k}} ^{\prime} \widetilde{v}_{j, \textbf{k}}^{x} \widetilde{v}_{\bar{j}, \textbf{Q+q}-\textbf{k}}^{x} G_{j, \textbf{k}}^{R} G_{j, \textbf{k}}^{A} G_{\bar{j}, \textbf{Q+q}-\textbf{k}}^{R} G_{\bar{j}, \textbf{Q+q}-\textbf{k}}^{A} \\
=&- \frac{2 \pi \eta_{j}\eta_{\bar{j}}v_{j}v_{\bar{j}}N\tau_{j}\tau_{\bar{j}}\tau}{\hbar^{3}} tanh(\lambda \frac {k_{min}}{Q}).
\end{aligned}
\end{equation}
In the limit $\textbf{q} \rightarrow 0$. The condition $\epsilon_{j,\textbf{k}}=\epsilon_{\bar{j},\textbf{Q}-\textbf{k}}$ is used to obtain Eq. (\ref{eq36}). Next, we need to calculate the summation of Cooperon. By using the explicit expressions of the Cooperons given in Eqs. (\ref{eq22}) and (\ref{eq32}), and transforming the summation over $\textbf{q}$ into an integral, the conductivity correction can be written in the following form \cite{datta1997electronic,shan2012spin}:

\begin{equation}\label{eq37}
    \begin{aligned}
\it \Delta \sigma_{intra}&=-\sum_{j}\frac{\eta_{j}^{2}}{2 \pi} \frac{e^{2}}{h} \sum_{L=0,1,2} (-1)^{L} \int_{l_{\varphi}^{-2}}^{l_{j, e}^{-2}} \frac{1}{l_{j, L}^{-2}+q^{2}} d q^{2} \\
&=-\sum_{j}\frac{\eta_{j}^{2}}{2 \pi} \frac{e^{2}}{h} \sum_{L=0,1,2} (-1)^{L}  \ln \frac{l_{j, L}^{-2}+l_{j, e}^{-2}}{l_{j, L}^{-2}+l_{\varphi}^{-2}}.
\end{aligned}
\end{equation}

\begin{equation}\label{eq38}
\begin{aligned}
\it \Delta \sigma_{inter}=&-\sum_{j}\frac{\eta_{j} \eta_{\bar{j}}}{2 \pi} \frac{e^{2}}{h} \frac{\tau_{j}\tau_{\bar{j}}}{\tau^{2}} \frac{v_{j}v_{\bar{j}}}{v^{2}} \sum_{L=0,1,2} \\
& \times (-1)^{L} \alpha_{t,L} \int_{l_{\varphi}^{-2}}^{l_{t, e}^{-2}} \frac{1}{l_{t, L}^{-2}+q^{2}} d q^{2} \\
=&-\sum_{j}\frac{\eta_{j} \eta_{\bar{j}}}{2 \pi} \frac{e^{2}}{h} \frac{\tau_{j}\tau_{\bar{j}}}{\tau^{2}} \frac{v_{j}v_{\bar{j}}}{v^{2}} \sum_{L=0,1,2} \\
& \times (-1)^{L} \alpha_{t,L} \ln \frac{l_{t, L}^{-2}+l_{t, e}^{-2}}{l_{t, L}^{-2}+l_{\varphi}^{-2}}.
\end{aligned}
\end{equation}
Here, $l_{j, e}$ represents the mean free path of valley $j$, where it's inverse square is taken as the upper limit of the integral with respect to $\textbf{q}^2$. For inter-valley scattering, the effective mean free path is denoted as $l_{t,e}$ and is calculated as $l_{t,e}=\sqrt{D\tau}$. The coherence length $l_{\varphi}$ has a strong temperature dependence which is given by $l_{\varphi} \propto T^{-1/2}$. We provide an explicit expression for coherence length as $l_{\varphi} = l_{1,e}(T/T^{*})^{-1/2} $, where $T^{*}$ is the temperature at which coherence length equals the mean free length of the electron in the valley $\textbf{K}_{1}$. The coefficient $\alpha_{t,L}$ has been given in Eq. (\ref{eq32}). The total conductivity correction is the sum of the contributions from intra- and inter-scattering:

\begin{equation}\label{eq39}
\it \Delta \sigma=\it \Delta \sigma_{intra}+\it \Delta \sigma_{inter}.
\end{equation}
We investigate the conductivity correction of the system with twist angle equals $3^{\circ}$, the Fermi energy $E_{F} = 42$ meV, the gap of valley $\textbf{K}_2$ $\it \Delta$ = 53 meV, $\lambda$ = 0.1, $\beta = \sqrt{3/4} \it \Delta K_{\theta = 1.65^{\circ}}$\cite{phinney2021strong}, where $\it \Delta K_{\theta = 1.65^{\circ}}$ is the wave vector difference between two Dirac points for a twist angle of $1.65^{\circ}$. The results are presented in Figs. \ref{figure 3}(a) and \ref{figure 3}(b). In Fig. \ref{figure 3}(a), the Fermi surface areas of the two valleys are unequal when the bias voltage is turned off. We observe that the contribution from inter-valley scattering denoted by the green line is negligible and the total correction is negative across the entire temperature range, indicating a weak localization regime. When the bias voltage is turned on and the energy shift $U$ = -8 meV, the bottom energy band dropped and the Fermi surfaces of the two valleys become similar in size. In Fig. \ref{figure 3}(b), we find that inter-valley scattering plays a significant role at this point, and the sign of the total conductivity correction changes to a positive value at low temperatures, indicating a crossover from weak localization to weak anti-localization by tuning the bias voltage.

Next, we turn to the calculation of magneto-conductivity. The external vertical magnetic field induces the appearance of Landau levels, which discretizes the wave vector \textbf{q} into a series of discrete numbers: $q_{n}=\sqrt{(n+1 / 2) / l_{B}^{2}}$, where $l_{B}$ is the magnetic length of Cooperon defined as $l_{B}=\sqrt{\hbar / 4 e \left | B \right | }$\cite{datta1997electronic}. The summation over wave vector $\textbf{q}$ can then be transformed into a summation over the energy level index $n$. The finite-field conductivity correction is expressed in terms of digamma functions $\Psi$,

\begin{figure*}
    \centering
    \includegraphics[width=16cm,height=9cm]{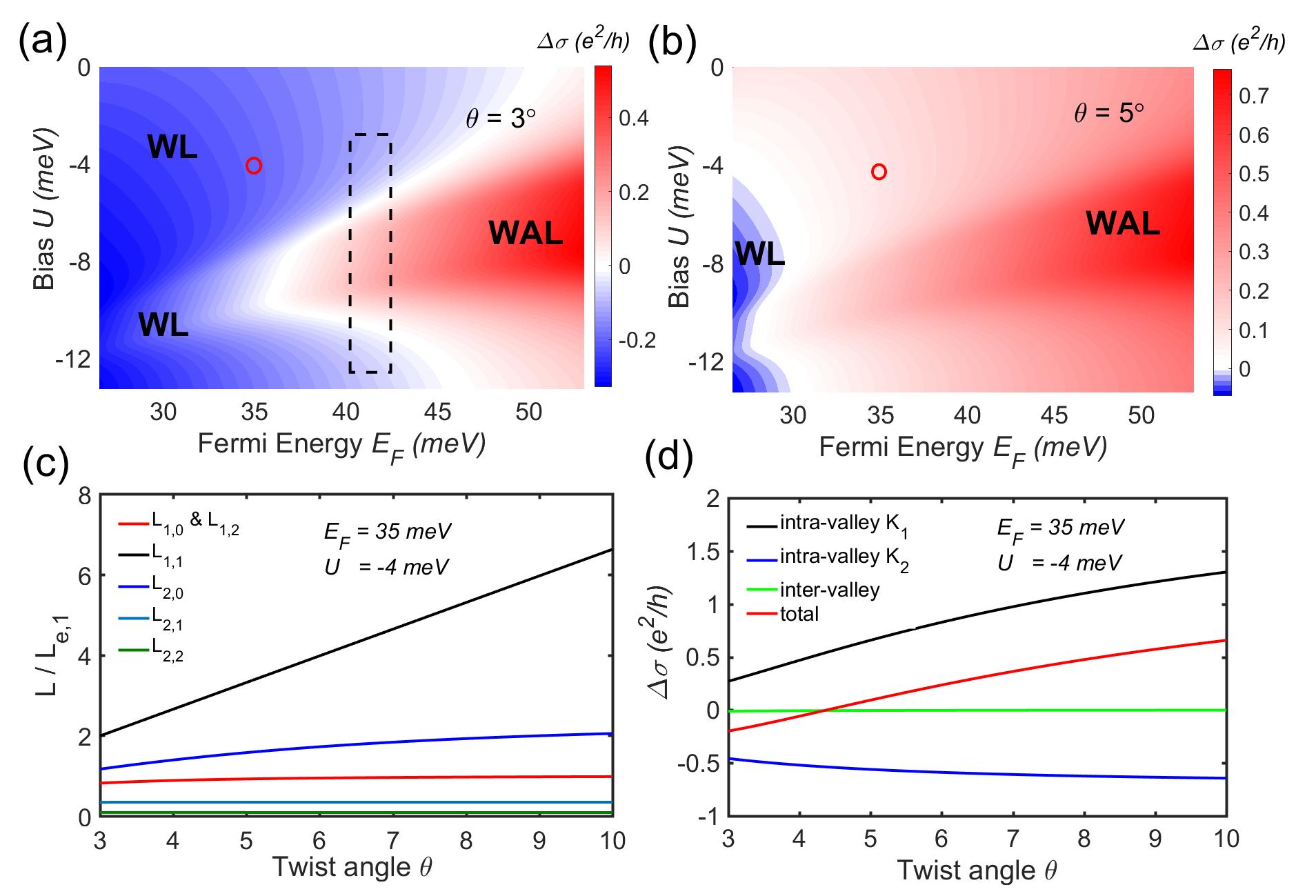}
    \caption{(a) Phase diagram of the system with a twist angle of $\theta = 3^{\circ}$. The dotted box marks the double crossover region. (b) Phase diagram of the system with a twist angle of $\theta = 5^{\circ}$. (c) The relation between the length of the six intra-valley scattering channels and twist angle. (d) Conductivity correction versus twist angle of the system. The Fermi energy $E_{F}$ and $U$ in (c) and (d) are corresponding to the red circle in (a) and (b). The total conductivity correction changes from negative to positive when the twist angle is around $4.5^\circ$. The temperature is set to be $T/T^{*}$ = 0.01.}
    \label{figure 4}
\end{figure*}

\begin{equation}\label{eq40}
\begin{aligned}
\it \Delta \sigma_{intra}(B)=&-\sum_{j, L} \frac{\eta^{2}}{2 \pi} \frac{e^{2}}{h} (-1)^{L} \{ \Psi(\frac{1}{2}+\frac{l_{B}^{2}}{l_{j, L}^2}+\frac{l_{B}^2}{l_{j,e}^2}). \\
&-\Psi(\frac{1}{2}+\frac{l_{B}^{2}}{l_{j, L}^2}+\frac{l_{B}^2}{l_{\phi}^2})\}.
\end{aligned}
\end{equation}

\begin{equation}\label{eq41}
\begin{aligned}
\it \Delta \sigma_{inter}(B)=&-\sum_{j, L} \frac{\eta_{j} \eta_{\bar{j}}}{2 \pi} \frac{e^{2}}{h} \frac{\tau_{j}\tau_{\bar{j}}}{\tau^{2}} \frac{v_{j}v_{\bar{j}}}{v^{2}} (-1)^{L} \\
& \times \{ \Psi(\frac{1}{2}+\frac{l_{B}^{2}}{l_{t, L}^2}+\frac{l_{B}^2}{l_{t,e}^2})
-\Psi(\frac{1}{2}+\frac{l_{B}^{2}}{l_{t, L}^2}+\frac{l_{B}^2}{l_{\phi}^2})\}.
\end{aligned}
\end{equation}
The magneto-conductivity is defined as the change of the conductivity correction under a finite magnetic field:

\begin{equation}\label{eq42}
\it \Delta \sigma_{B}=\it \Delta \sigma(B)-\it \Delta \sigma(0).
\end{equation}\\
Figs. \ref{figure 3}(c) and \ref{figure 3}(d) present the magneto-conductivity curves. In the presence of magnetic field, time reversal symmetry is broken, suppressing quantum interference in the system. The total magneto-conductivity curve shown in Fig. \ref{figure 3}(c) exhibits an upward trend upon increasing the magnetic field beyond a certain threshold, suggesting the existence of the original weak localization state. On the other hand, the curve in Fig. \ref{figure 3}(d) demonstrates an obvious downward trend with an increasing magnetic field, indicating the system's original state is weak antilocalization. Our results provide clear evidence of the WL-WAL crossover via tuning of the bias voltage, and experimental measurement of the magneto-conductivity is feasible to verify it. \\

Furthermore, we present phase diagrams of WL and WAL in the Fermi energy $E_{F}$ and bias voltage $U$ plane for different twist angles in Figs. \ref{figure 4}(a) and \ref{figure 4}(b).  The system with a small twist angle corresponds to stronger inter-valley scattering, as evidenced by Eq. (\ref{eq7}). In Fig. \ref{figure 4}(a), for the system with a twist angle of $\theta = 3^{\circ}$, we observe that the red WAL region penetrates deep into the blue WL region, resulting in a double crossover of WL-WAL-WL. This area is represented by the dotted box, with the Fermi energy ranging from 40 meV to 43 meV. This double crossover can be elucidated by the connection between inter-valley scattering and the bias voltage. As we tune the bias voltage, the energy band of valley $\textbf{K}_2$ shifts, and when the Fermi surface area of the two valleys approaches each other, inter-valley scattering dramatically contributes to the conductivity correction. As one continues to adjust the bias voltage, the difference in the Fermi surface size of the two valleys increases. The contribution of inter-valley scattering to the conductivity correction becomes very weak. In Fig. \ref{figure 4}(b), when the twist angle is $\theta = 5^{\circ}$, in a sharp contrast to Fig. 4(a), the WAL and WL regions are clearly separated and the double crossover disappears. We conclude that inter-valley scattering plays a critical role in the occurrence of double crossover. From Fig. 4(b), we notice that the conditions for detecting the double crossover are quite stringent. The twist angle of the system cannot be large. As in Fig. 4(b), the original WL region turn into a WAL region when the twist angle increases to $5^\circ$, and the double crossover fades away. In Fig. 4(c), we investigate the parameter $E_{F}$ = 35 meV and $U$ = -4 meV that corresponds to the red circle in Fig. 4(a) and 4(b). We find that the increase in twist angle mainly affects the dominant intra-valley scattering length of the gapless cone $l_{1,1}$, which has the fastest rate of increase among all the six intra-valley scattering channels. The dominant intra-valley scattering length of the gapped cone $l_{2,0}$ does not change dramatically with a varying twisted angle. This leads to a rapid increase in the gapless valley contribution to WAL. Consequently, the value of WAL can surpass WL at certain twist angle, as demonstrated in Fig. 4(d). This explains why WAL occupies the majority of the phase diagram in Fig. 4(b) when the twist angle increases.

\section{Summary and discussions}\label{V}
To summarize, we investigated the quantum conductivity corrections for weak (anti-)localization effects in the system with adjacent gapless and gapped Dirac cones. Our findings reveals that inter-valley scattering has a significant impact on conductivity correction, particularly when the Fermi surface of the two valleys are equal in size. Through standard calculation, we observe a double crossover in a specific range of Fermi energy by tuning bias voltage when the twist angle is small.  Our empirical method for handling inter-valley scattering with unequal Fermi surface areas may be applicable to other multi-valley systems, and our calculations may also be beneficial for future experimental investigations on the quantum transport properties of two-dimensional twisted moir\'e materials.

\section{Acknowledgements}\label{VI}
This work was financially supported by the National Basic Research Program of China (Grants No. 2017YFA0303301), the National Natural Science Foundation of China (Grants No. 12022407).

\bibliography{reference}

\begin{thebibliography}{46}%
\makeatletter
\providecommand \@ifxundefined [1]{%
 \@ifx{#1\undefined}
}%
\providecommand \@ifnum [1]{%
 \ifnum #1\expandafter \@firstoftwo
 \else \expandafter \@secondoftwo
 \fi
}%
\providecommand \@ifx [1]{%
 \ifx #1\expandafter \@firstoftwo
 \else \expandafter \@secondoftwo
 \fi
}%
\providecommand \natexlab [1]{#1}%
\providecommand \enquote  [1]{``#1''}%
\providecommand \bibnamefont  [1]{#1}%
\providecommand \bibfnamefont [1]{#1}%
\providecommand \citenamefont [1]{#1}%
\providecommand \href@noop [0]{\@secondoftwo}%
\providecommand \href [0]{\begingroup \@sanitize@url \@href}%
\providecommand \@href[1]{\@@startlink{#1}\@@href}%
\providecommand \@@href[1]{\endgroup#1\@@endlink}%
\providecommand \@sanitize@url [0]{\catcode `\\12\catcode `\$12\catcode
  `\&12\catcode `\#12\catcode `\^12\catcode `\_12\catcode `\%12\relax}%
\providecommand \@@startlink[1]{}%
\providecommand \@@endlink[0]{}%
\providecommand \url  [0]{\begingroup\@sanitize@url \@url }%
\providecommand \@url [1]{\endgroup\@href {#1}{\urlprefix }}%
\providecommand \urlprefix  [0]{URL }%
\providecommand \Eprint [0]{\href }%
\providecommand \doibase [0]{http://dx.doi.org/}%
\providecommand \selectlanguage [0]{\@gobble}%
\providecommand \bibinfo  [0]{\@secondoftwo}%
\providecommand \bibfield  [0]{\@secondoftwo}%
\providecommand \translation [1]{[#1]}%
\providecommand \BibitemOpen [0]{}%
\providecommand \bibitemStop [0]{}%
\providecommand \bibitemNoStop [0]{.\EOS\space}%
\providecommand \EOS [0]{\spacefactor3000\relax}%
\providecommand \BibitemShut  [1]{\csname bibitem#1\endcsname}%
\let\auto@bib@innerbib\@empty
\bibitem [{\citenamefont {Hikami}\ \emph {et~al.}(1980)\citenamefont {Hikami},
  \citenamefont {Larkin},\ and\ \citenamefont {Nagaoka}}]{hikami1980spin}%
  \BibitemOpen
  \bibfield  {author} {\bibinfo {author} {\bibfnamefont {S.}~\bibnamefont
  {Hikami}}, \bibinfo {author} {\bibfnamefont {A.~I.}\ \bibnamefont {Larkin}},
  \ and\ \bibinfo {author} {\bibfnamefont {Y.}~\bibnamefont {Nagaoka}},\ }\href
  {\doibase 10.1143/PTP.63.707} {\bibfield  {journal} {\bibinfo  {journal}
  {Progress of Theoretical Physics}\ }\textbf {\bibinfo {volume} {63}},\
  \bibinfo {pages} {707} (\bibinfo {year} {1980})}\BibitemShut {NoStop}%
\bibitem [{\citenamefont {Altshuler}\ \emph {et~al.}(1980)\citenamefont
  {Altshuler}, \citenamefont {Khmel'nitzkii}, \citenamefont {Larkin},\ and\
  \citenamefont {Lee}}]{altshuler1980magnetoresistance}%
  \BibitemOpen
  \bibfield  {author} {\bibinfo {author} {\bibfnamefont {B.~L.}\ \bibnamefont
  {Altshuler}}, \bibinfo {author} {\bibfnamefont {D.}~\bibnamefont
  {Khmel'nitzkii}}, \bibinfo {author} {\bibfnamefont {A.~I.}\ \bibnamefont
  {Larkin}}, \ and\ \bibinfo {author} {\bibfnamefont {P.~A.}\ \bibnamefont
  {Lee}},\ }\href {\doibase 10.1103/PhysRevB.22.5142} {\bibfield  {journal}
  {\bibinfo  {journal} {Phys. Rev. B}\ }\textbf {\bibinfo {volume} {22}},\
  \bibinfo {pages} {5142} (\bibinfo {year} {1980})}\BibitemShut {NoStop}%
\bibitem [{\citenamefont {Novoselov}\ \emph {et~al.}(2004)\citenamefont
  {Novoselov}, \citenamefont {Geim}, \citenamefont {Morozov}, \citenamefont
  {Jiang}, \citenamefont {Zhang}, \citenamefont {Dubonos}, \citenamefont
  {Grigorieva},\ and\ \citenamefont {Firsov}}]{novoselov2004electric}%
  \BibitemOpen
  \bibfield  {author} {\bibinfo {author} {\bibfnamefont {K.~S.}\ \bibnamefont
  {Novoselov}}, \bibinfo {author} {\bibfnamefont {A.~K.}\ \bibnamefont {Geim}},
  \bibinfo {author} {\bibfnamefont {S.~V.}\ \bibnamefont {Morozov}}, \bibinfo
  {author} {\bibfnamefont {D.}~\bibnamefont {Jiang}}, \bibinfo {author}
  {\bibfnamefont {Y.}~\bibnamefont {Zhang}}, \bibinfo {author} {\bibfnamefont
  {S.~V.}\ \bibnamefont {Dubonos}}, \bibinfo {author} {\bibfnamefont {I.~V.}\
  \bibnamefont {Grigorieva}}, \ and\ \bibinfo {author} {\bibfnamefont {A.~A.}\
  \bibnamefont {Firsov}},\ }\href {\doibase 10.1126/science.1102896} {\bibfield
   {journal} {\bibinfo  {journal} {Science}\ }\textbf {\bibinfo {volume}
  {306}},\ \bibinfo {pages} {666} (\bibinfo {year} {2004})}\BibitemShut
  {NoStop}%
\bibitem [{\citenamefont {Ando}\ \emph {et~al.}(1998)\citenamefont {Ando},
  \citenamefont {Nakanishi},\ and\ \citenamefont {Saito}}]{ando1998berry}%
  \BibitemOpen
  \bibfield  {author} {\bibinfo {author} {\bibfnamefont {T.}~\bibnamefont
  {Ando}}, \bibinfo {author} {\bibfnamefont {T.}~\bibnamefont {Nakanishi}}, \
  and\ \bibinfo {author} {\bibfnamefont {R.}~\bibnamefont {Saito}},\ }\href
  {\doibase 10.1143/JPSJ.67.2857} {\bibfield  {journal} {\bibinfo  {journal}
  {Journal of the Physical Society of Japan}\ }\textbf {\bibinfo {volume}
  {67}},\ \bibinfo {pages} {2857} (\bibinfo {year} {1998})}\BibitemShut
  {NoStop}%
\bibitem [{\citenamefont {Suzuura}\ and\ \citenamefont
  {Ando}(2002)}]{suzuura2002crossover}%
  \BibitemOpen
  \bibfield  {author} {\bibinfo {author} {\bibfnamefont {H.}~\bibnamefont
  {Suzuura}}\ and\ \bibinfo {author} {\bibfnamefont {T.}~\bibnamefont {Ando}},\
  }\href {\doibase 10.1103/PhysRevLett.89.266603} {\bibfield  {journal}
  {\bibinfo  {journal} {Phys. Rev. Lett.}\ }\textbf {\bibinfo {volume} {89}},\
  \bibinfo {pages} {266603} (\bibinfo {year} {2002})}\BibitemShut {NoStop}%
\bibitem [{\citenamefont {McCann}\ \emph {et~al.}(2006)\citenamefont {McCann},
  \citenamefont {Kechedzhi}, \citenamefont {Fal'ko}, \citenamefont {Suzuura},
  \citenamefont {Ando},\ and\ \citenamefont {Altshuler}}]{mccann2006weak}%
  \BibitemOpen
  \bibfield  {author} {\bibinfo {author} {\bibfnamefont {E.}~\bibnamefont
  {McCann}}, \bibinfo {author} {\bibfnamefont {K.}~\bibnamefont {Kechedzhi}},
  \bibinfo {author} {\bibfnamefont {V.~I.}\ \bibnamefont {Fal'ko}}, \bibinfo
  {author} {\bibfnamefont {H.}~\bibnamefont {Suzuura}}, \bibinfo {author}
  {\bibfnamefont {T.}~\bibnamefont {Ando}}, \ and\ \bibinfo {author}
  {\bibfnamefont {B.~L.}\ \bibnamefont {Altshuler}},\ }\href {\doibase
  10.1103/PhysRevLett.97.146805} {\bibfield  {journal} {\bibinfo  {journal}
  {Phys. Rev. Lett.}\ }\textbf {\bibinfo {volume} {97}},\ \bibinfo {pages}
  {146805} (\bibinfo {year} {2006})}\BibitemShut {NoStop}%
\bibitem [{\citenamefont {Morozov}\ \emph {et~al.}(2006)\citenamefont
  {Morozov}, \citenamefont {Novoselov}, \citenamefont {Katsnelson},
  \citenamefont {Schedin}, \citenamefont {Ponomarenko}, \citenamefont {Jiang},\
  and\ \citenamefont {Geim}}]{morozov2006strong}%
  \BibitemOpen
  \bibfield  {author} {\bibinfo {author} {\bibfnamefont {S.~V.}\ \bibnamefont
  {Morozov}}, \bibinfo {author} {\bibfnamefont {K.~S.}\ \bibnamefont
  {Novoselov}}, \bibinfo {author} {\bibfnamefont {M.~I.}\ \bibnamefont
  {Katsnelson}}, \bibinfo {author} {\bibfnamefont {F.}~\bibnamefont {Schedin}},
  \bibinfo {author} {\bibfnamefont {L.~A.}\ \bibnamefont {Ponomarenko}},
  \bibinfo {author} {\bibfnamefont {D.}~\bibnamefont {Jiang}}, \ and\ \bibinfo
  {author} {\bibfnamefont {A.~K.}\ \bibnamefont {Geim}},\ }\href {\doibase
  10.1103/PhysRevLett.97.016801} {\bibfield  {journal} {\bibinfo  {journal}
  {Phys. Rev. Lett.}\ }\textbf {\bibinfo {volume} {97}},\ \bibinfo {pages}
  {016801} (\bibinfo {year} {2006})}\BibitemShut {NoStop}%
\bibitem [{\citenamefont {Wu}\ \emph {et~al.}(2007)\citenamefont {Wu},
  \citenamefont {Li}, \citenamefont {Song}, \citenamefont {Berger},\ and\
  \citenamefont {de~Heer}}]{wu2007weak}%
  \BibitemOpen
  \bibfield  {author} {\bibinfo {author} {\bibfnamefont {X.}~\bibnamefont
  {Wu}}, \bibinfo {author} {\bibfnamefont {X.}~\bibnamefont {Li}}, \bibinfo
  {author} {\bibfnamefont {Z.}~\bibnamefont {Song}}, \bibinfo {author}
  {\bibfnamefont {C.}~\bibnamefont {Berger}}, \ and\ \bibinfo {author}
  {\bibfnamefont {W.~A.}\ \bibnamefont {de~Heer}},\ }\href {\doibase
  10.1103/PhysRevLett.98.136801} {\bibfield  {journal} {\bibinfo  {journal}
  {Phys. Rev. Lett.}\ }\textbf {\bibinfo {volume} {98}},\ \bibinfo {pages}
  {136801} (\bibinfo {year} {2007})}\BibitemShut {NoStop}%
\bibitem [{\citenamefont {Kechedzhi}\ \emph
  {et~al.}(2007{\natexlab{a}})\citenamefont {Kechedzhi}, \citenamefont
  {McCann}, \citenamefont {Fal'ko}, \citenamefont {Suzuura}, \citenamefont
  {Ando},\ and\ \citenamefont {Altshuler}}]{kechedzhi2007weak}%
  \BibitemOpen
  \bibfield  {author} {\bibinfo {author} {\bibfnamefont {K.}~\bibnamefont
  {Kechedzhi}}, \bibinfo {author} {\bibfnamefont {E.}~\bibnamefont {McCann}},
  \bibinfo {author} {\bibfnamefont {V.~I.}\ \bibnamefont {Fal'ko}}, \bibinfo
  {author} {\bibfnamefont {H.}~\bibnamefont {Suzuura}}, \bibinfo {author}
  {\bibfnamefont {T.}~\bibnamefont {Ando}}, \ and\ \bibinfo {author}
  {\bibfnamefont {B.~L.}\ \bibnamefont {Altshuler}},\ }\href {\doibase
  10.1140/epjst/e2007-00224-6} {\bibfield  {journal} {\bibinfo  {journal} {The
  European Physical Journal Special Topics}\ }\textbf {\bibinfo {volume}
  {148}},\ \bibinfo {pages} {39} (\bibinfo {year}
  {2007}{\natexlab{a}})}\BibitemShut {NoStop}%
\bibitem [{\citenamefont {Fal'ko}\ \emph {et~al.}(2007)\citenamefont {Fal'ko},
  \citenamefont {Kechedzhi}, \citenamefont {McCann}, \citenamefont {Altshuler},
  \citenamefont {Suzuura},\ and\ \citenamefont {Ando}}]{fal2007weak}%
  \BibitemOpen
  \bibfield  {author} {\bibinfo {author} {\bibfnamefont {V.~I.}\ \bibnamefont
  {Fal'ko}}, \bibinfo {author} {\bibfnamefont {K.}~\bibnamefont {Kechedzhi}},
  \bibinfo {author} {\bibfnamefont {E.}~\bibnamefont {McCann}}, \bibinfo
  {author} {\bibfnamefont {B.}~\bibnamefont {Altshuler}}, \bibinfo {author}
  {\bibfnamefont {H.}~\bibnamefont {Suzuura}}, \ and\ \bibinfo {author}
  {\bibfnamefont {T.}~\bibnamefont {Ando}},\ }\href {\doibase
  https://doi.org/10.1016/j.ssc.2007.03.049} {\bibfield  {journal} {\bibinfo
  {journal} {Solid State Communications}\ }\textbf {\bibinfo {volume} {143}},\
  \bibinfo {pages} {33} (\bibinfo {year} {2007})},\ \bibinfo {note} {exploring
  graphene}\BibitemShut {NoStop}%
\bibitem [{\citenamefont {Yan}\ and\ \citenamefont {Ting}(2008)}]{yan2008weak}%
  \BibitemOpen
  \bibfield  {author} {\bibinfo {author} {\bibfnamefont {X.-Z.}\ \bibnamefont
  {Yan}}\ and\ \bibinfo {author} {\bibfnamefont {C.~S.}\ \bibnamefont {Ting}},\
  }\href {\doibase 10.1103/PhysRevLett.101.126801} {\bibfield  {journal}
  {\bibinfo  {journal} {Phys. Rev. Lett.}\ }\textbf {\bibinfo {volume} {101}},\
  \bibinfo {pages} {126801} (\bibinfo {year} {2008})}\BibitemShut {NoStop}%
\bibitem [{\citenamefont {Tikhonenko}\ \emph {et~al.}(2008)\citenamefont
  {Tikhonenko}, \citenamefont {Horsell}, \citenamefont {Gorbachev},\ and\
  \citenamefont {Savchenko}}]{tikhonenko2008weak}%
  \BibitemOpen
  \bibfield  {author} {\bibinfo {author} {\bibfnamefont {F.~V.}\ \bibnamefont
  {Tikhonenko}}, \bibinfo {author} {\bibfnamefont {D.~W.}\ \bibnamefont
  {Horsell}}, \bibinfo {author} {\bibfnamefont {R.~V.}\ \bibnamefont
  {Gorbachev}}, \ and\ \bibinfo {author} {\bibfnamefont {A.~K.}\ \bibnamefont
  {Savchenko}},\ }\href {\doibase 10.1103/PhysRevLett.100.056802} {\bibfield
  {journal} {\bibinfo  {journal} {Phys. Rev. Lett.}\ }\textbf {\bibinfo
  {volume} {100}},\ \bibinfo {pages} {056802} (\bibinfo {year}
  {2008})}\BibitemShut {NoStop}%
\bibitem [{\citenamefont {Tikhonenko}\ \emph {et~al.}(2009)\citenamefont
  {Tikhonenko}, \citenamefont {Kozikov}, \citenamefont {Savchenko},\ and\
  \citenamefont {Gorbachev}}]{tikhonenko2009transition}%
  \BibitemOpen
  \bibfield  {author} {\bibinfo {author} {\bibfnamefont {F.~V.}\ \bibnamefont
  {Tikhonenko}}, \bibinfo {author} {\bibfnamefont {A.~A.}\ \bibnamefont
  {Kozikov}}, \bibinfo {author} {\bibfnamefont {A.~K.}\ \bibnamefont
  {Savchenko}}, \ and\ \bibinfo {author} {\bibfnamefont {R.~V.}\ \bibnamefont
  {Gorbachev}},\ }\href {\doibase 10.1103/PhysRevLett.103.226801} {\bibfield
  {journal} {\bibinfo  {journal} {Phys. Rev. Lett.}\ }\textbf {\bibinfo
  {volume} {103}},\ \bibinfo {pages} {226801} (\bibinfo {year}
  {2009})}\BibitemShut {NoStop}%
\bibitem [{\citenamefont {Hilke}\ \emph {et~al.}(2014)\citenamefont {Hilke},
  \citenamefont {Massicotte}, \citenamefont {Whiteway},\ and\ \citenamefont
  {Yu}}]{hilke2014weak}%
  \BibitemOpen
  \bibfield  {author} {\bibinfo {author} {\bibfnamefont {M.}~\bibnamefont
  {Hilke}}, \bibinfo {author} {\bibfnamefont {M.}~\bibnamefont {Massicotte}},
  \bibinfo {author} {\bibfnamefont {E.}~\bibnamefont {Whiteway}}, \ and\
  \bibinfo {author} {\bibfnamefont {V.}~\bibnamefont {Yu}},\ }\href {\doibase
  10.1155/2014/737296} {\bibfield  {journal} {\bibinfo  {journal} {The
  Scientific World Journal}\ }\textbf {\bibinfo {volume} {2014}},\ \bibinfo
  {pages} {737296} (\bibinfo {year} {2014})}\BibitemShut {NoStop}%
\bibitem [{\citenamefont {Giovannetti}\ \emph {et~al.}(2007)\citenamefont
  {Giovannetti}, \citenamefont {Khomyakov}, \citenamefont {Brocks},
  \citenamefont {Kelly},\ and\ \citenamefont {van~den
  Brink}}]{giovannetti2007substrate}%
  \BibitemOpen
  \bibfield  {author} {\bibinfo {author} {\bibfnamefont {G.}~\bibnamefont
  {Giovannetti}}, \bibinfo {author} {\bibfnamefont {P.~A.}\ \bibnamefont
  {Khomyakov}}, \bibinfo {author} {\bibfnamefont {G.}~\bibnamefont {Brocks}},
  \bibinfo {author} {\bibfnamefont {P.~J.}\ \bibnamefont {Kelly}}, \ and\
  \bibinfo {author} {\bibfnamefont {J.}~\bibnamefont {van~den Brink}},\ }\href
  {\doibase 10.1103/PhysRevB.76.073103} {\bibfield  {journal} {\bibinfo
  {journal} {Phys. Rev. B}\ }\textbf {\bibinfo {volume} {76}},\ \bibinfo
  {pages} {073103} (\bibinfo {year} {2007})}\BibitemShut {NoStop}%
\bibitem [{\citenamefont {Kechedzhi}\ \emph
  {et~al.}(2007{\natexlab{b}})\citenamefont {Kechedzhi}, \citenamefont
  {Fal'ko}, \citenamefont {McCann},\ and\ \citenamefont
  {Altshuler}}]{kechedzhi2007influence}%
  \BibitemOpen
  \bibfield  {author} {\bibinfo {author} {\bibfnamefont {K.}~\bibnamefont
  {Kechedzhi}}, \bibinfo {author} {\bibfnamefont {V.~I.}\ \bibnamefont
  {Fal'ko}}, \bibinfo {author} {\bibfnamefont {E.}~\bibnamefont {McCann}}, \
  and\ \bibinfo {author} {\bibfnamefont {B.~L.}\ \bibnamefont {Altshuler}},\
  }\href {\doibase 10.1103/PhysRevLett.98.176806} {\bibfield  {journal}
  {\bibinfo  {journal} {Phys. Rev. Lett.}\ }\textbf {\bibinfo {volume} {98}},\
  \bibinfo {pages} {176806} (\bibinfo {year} {2007}{\natexlab{b}})}\BibitemShut
  {NoStop}%
\bibitem [{\citenamefont {Horsell}\ \emph {et~al.}(2008)\citenamefont
  {Horsell}, \citenamefont {Tikhonenko}, \citenamefont {Gorbachev},\ and\
  \citenamefont {Savchenko}}]{horsell2008weak}%
  \BibitemOpen
  \bibfield  {author} {\bibinfo {author} {\bibfnamefont {D.}~\bibnamefont
  {Horsell}}, \bibinfo {author} {\bibfnamefont {F.}~\bibnamefont {Tikhonenko}},
  \bibinfo {author} {\bibfnamefont {R.}~\bibnamefont {Gorbachev}}, \ and\
  \bibinfo {author} {\bibfnamefont {A.}~\bibnamefont {Savchenko}},\ }\href
  {\doibase 10.1098/rsta.2007.2159} {\bibfield  {journal} {\bibinfo  {journal}
  {Philos. Trans. R. Soc., A}\ }\textbf {\bibinfo {volume} {366}},\ \bibinfo
  {pages} {245} (\bibinfo {year} {2008})}\BibitemShut {NoStop}%
\bibitem [{\citenamefont {Liao}\ \emph {et~al.}(2010)\citenamefont {Liao},
  \citenamefont {Han}, \citenamefont {Wu},\ and\ \citenamefont
  {Yu}}]{liao2010gate}%
  \BibitemOpen
  \bibfield  {author} {\bibinfo {author} {\bibfnamefont {Z.-M.}\ \bibnamefont
  {Liao}}, \bibinfo {author} {\bibfnamefont {B.-H.}\ \bibnamefont {Han}},
  \bibinfo {author} {\bibfnamefont {H.-C.}\ \bibnamefont {Wu}}, \ and\ \bibinfo
  {author} {\bibfnamefont {D.-P.}\ \bibnamefont {Yu}},\ }\href {\doibase
  10.1063/1.3505310} {\bibfield  {journal} {\bibinfo  {journal} {Applied
  Physics Letters}\ }\textbf {\bibinfo {volume} {97}},\ \bibinfo {pages}
  {163110} (\bibinfo {year} {2010})}\BibitemShut {NoStop}%
\bibitem [{\citenamefont {Amann}\ \emph {et~al.}(2022)\citenamefont {Amann},
  \citenamefont {V\"olkl}, \citenamefont {Rockinger}, \citenamefont {Kochan},
  \citenamefont {Watanabe}, \citenamefont {Taniguchi}, \citenamefont {Fabian},
  \citenamefont {Weiss},\ and\ \citenamefont
  {Eroms}}]{amann2022counterintuitive}%
  \BibitemOpen
  \bibfield  {author} {\bibinfo {author} {\bibfnamefont {J.}~\bibnamefont
  {Amann}}, \bibinfo {author} {\bibfnamefont {T.}~\bibnamefont {V\"olkl}},
  \bibinfo {author} {\bibfnamefont {T.}~\bibnamefont {Rockinger}}, \bibinfo
  {author} {\bibfnamefont {D.}~\bibnamefont {Kochan}}, \bibinfo {author}
  {\bibfnamefont {K.}~\bibnamefont {Watanabe}}, \bibinfo {author}
  {\bibfnamefont {T.}~\bibnamefont {Taniguchi}}, \bibinfo {author}
  {\bibfnamefont {J.}~\bibnamefont {Fabian}}, \bibinfo {author} {\bibfnamefont
  {D.}~\bibnamefont {Weiss}}, \ and\ \bibinfo {author} {\bibfnamefont
  {J.}~\bibnamefont {Eroms}},\ }\href {\doibase 10.1103/PhysRevB.105.115425}
  {\bibfield  {journal} {\bibinfo  {journal} {Phys. Rev. B}\ }\textbf {\bibinfo
  {volume} {105}},\ \bibinfo {pages} {115425} (\bibinfo {year}
  {2022})}\BibitemShut {NoStop}%
\bibitem [{\citenamefont {Wang}\ \emph {et~al.}(2015)\citenamefont {Wang},
  \citenamefont {Ki}, \citenamefont {Chen}, \citenamefont {Berger},
  \citenamefont {MacDonald},\ and\ \citenamefont {Morpurgo}}]{wang2015strong}%
  \BibitemOpen
  \bibfield  {author} {\bibinfo {author} {\bibfnamefont {Z.}~\bibnamefont
  {Wang}}, \bibinfo {author} {\bibfnamefont {D.-K.}\ \bibnamefont {Ki}},
  \bibinfo {author} {\bibfnamefont {H.}~\bibnamefont {Chen}}, \bibinfo {author}
  {\bibfnamefont {H.}~\bibnamefont {Berger}}, \bibinfo {author} {\bibfnamefont
  {A.~H.}\ \bibnamefont {MacDonald}}, \ and\ \bibinfo {author} {\bibfnamefont
  {A.~F.}\ \bibnamefont {Morpurgo}},\ }\href {\doibase 10.1038/ncomms9339}
  {\bibfield  {journal} {\bibinfo  {journal} {Nature Communications}\ }\textbf
  {\bibinfo {volume} {6}},\ \bibinfo {pages} {8339} (\bibinfo {year}
  {2015})}\BibitemShut {NoStop}%
\bibitem [{\citenamefont {Wang}\ \emph {et~al.}(2016)\citenamefont {Wang},
  \citenamefont {Ki}, \citenamefont {Khoo}, \citenamefont {Mauro},
  \citenamefont {Berger}, \citenamefont {Levitov},\ and\ \citenamefont
  {Morpurgo}}]{wang2016origin}%
  \BibitemOpen
  \bibfield  {author} {\bibinfo {author} {\bibfnamefont {Z.}~\bibnamefont
  {Wang}}, \bibinfo {author} {\bibfnamefont {D.-K.}\ \bibnamefont {Ki}},
  \bibinfo {author} {\bibfnamefont {J.~Y.}\ \bibnamefont {Khoo}}, \bibinfo
  {author} {\bibfnamefont {D.}~\bibnamefont {Mauro}}, \bibinfo {author}
  {\bibfnamefont {H.}~\bibnamefont {Berger}}, \bibinfo {author} {\bibfnamefont
  {L.~S.}\ \bibnamefont {Levitov}}, \ and\ \bibinfo {author} {\bibfnamefont
  {A.~F.}\ \bibnamefont {Morpurgo}},\ }\href {\doibase
  10.1103/PhysRevX.6.041020} {\bibfield  {journal} {\bibinfo  {journal} {Phys.
  Rev. X}\ }\textbf {\bibinfo {volume} {6}},\ \bibinfo {pages} {041020}
  (\bibinfo {year} {2016})}\BibitemShut {NoStop}%
\bibitem [{\citenamefont {Yang}\ \emph {et~al.}(2016)\citenamefont {Yang},
  \citenamefont {Tu}, \citenamefont {Kim}, \citenamefont {Wu}, \citenamefont
  {Wang}, \citenamefont {Alicea}, \citenamefont {Wu}, \citenamefont
  {Bockrath},\ and\ \citenamefont {Shi}}]{yang2016tunable}%
  \BibitemOpen
  \bibfield  {author} {\bibinfo {author} {\bibfnamefont {B.}~\bibnamefont
  {Yang}}, \bibinfo {author} {\bibfnamefont {M.-F.}\ \bibnamefont {Tu}},
  \bibinfo {author} {\bibfnamefont {J.}~\bibnamefont {Kim}}, \bibinfo {author}
  {\bibfnamefont {Y.}~\bibnamefont {Wu}}, \bibinfo {author} {\bibfnamefont
  {H.}~\bibnamefont {Wang}}, \bibinfo {author} {\bibfnamefont {J.}~\bibnamefont
  {Alicea}}, \bibinfo {author} {\bibfnamefont {R.}~\bibnamefont {Wu}}, \bibinfo
  {author} {\bibfnamefont {M.}~\bibnamefont {Bockrath}}, \ and\ \bibinfo
  {author} {\bibfnamefont {J.}~\bibnamefont {Shi}},\ }\href {\doibase
  10.1088/2053-1583/3/3/031012} {\bibfield  {journal} {\bibinfo  {journal} {2D
  Materials}\ }\textbf {\bibinfo {volume} {3}},\ \bibinfo {pages} {031012}
  (\bibinfo {year} {2016})}\BibitemShut {NoStop}%
\bibitem [{\citenamefont {V\"olkl}\ \emph {et~al.}(2017)\citenamefont
  {V\"olkl}, \citenamefont {Rockinger}, \citenamefont {Drienovsky},
  \citenamefont {Watanabe}, \citenamefont {Taniguchi}, \citenamefont {Weiss},\
  and\ \citenamefont {Eroms}}]{volkl2017magnetotransport}%
  \BibitemOpen
  \bibfield  {author} {\bibinfo {author} {\bibfnamefont {T.}~\bibnamefont
  {V\"olkl}}, \bibinfo {author} {\bibfnamefont {T.}~\bibnamefont {Rockinger}},
  \bibinfo {author} {\bibfnamefont {M.}~\bibnamefont {Drienovsky}}, \bibinfo
  {author} {\bibfnamefont {K.}~\bibnamefont {Watanabe}}, \bibinfo {author}
  {\bibfnamefont {T.}~\bibnamefont {Taniguchi}}, \bibinfo {author}
  {\bibfnamefont {D.}~\bibnamefont {Weiss}}, \ and\ \bibinfo {author}
  {\bibfnamefont {J.}~\bibnamefont {Eroms}},\ }\href {\doibase
  10.1103/PhysRevB.96.125405} {\bibfield  {journal} {\bibinfo  {journal} {Phys.
  Rev. B}\ }\textbf {\bibinfo {volume} {96}},\ \bibinfo {pages} {125405}
  (\bibinfo {year} {2017})}\BibitemShut {NoStop}%
\bibitem [{\citenamefont {Zihlmann}\ \emph {et~al.}(2018)\citenamefont
  {Zihlmann}, \citenamefont {Cummings}, \citenamefont {Garcia}, \citenamefont
  {Kedves}, \citenamefont {Watanabe}, \citenamefont {Taniguchi}, \citenamefont
  {Sch\"onenberger},\ and\ \citenamefont {Makk}}]{zihlmann2018large}%
  \BibitemOpen
  \bibfield  {author} {\bibinfo {author} {\bibfnamefont {S.}~\bibnamefont
  {Zihlmann}}, \bibinfo {author} {\bibfnamefont {A.~W.}\ \bibnamefont
  {Cummings}}, \bibinfo {author} {\bibfnamefont {J.~H.}\ \bibnamefont
  {Garcia}}, \bibinfo {author} {\bibfnamefont {M.}~\bibnamefont {Kedves}},
  \bibinfo {author} {\bibfnamefont {K.}~\bibnamefont {Watanabe}}, \bibinfo
  {author} {\bibfnamefont {T.}~\bibnamefont {Taniguchi}}, \bibinfo {author}
  {\bibfnamefont {C.}~\bibnamefont {Sch\"onenberger}}, \ and\ \bibinfo {author}
  {\bibfnamefont {P.}~\bibnamefont {Makk}},\ }\href {\doibase
  10.1103/PhysRevB.97.075434} {\bibfield  {journal} {\bibinfo  {journal} {Phys.
  Rev. B}\ }\textbf {\bibinfo {volume} {97}},\ \bibinfo {pages} {075434}
  (\bibinfo {year} {2018})}\BibitemShut {NoStop}%
\bibitem [{\citenamefont {Wakamura}\ \emph {et~al.}(2018)\citenamefont
  {Wakamura}, \citenamefont {Reale}, \citenamefont {Palczynski}, \citenamefont
  {Gu\'eron}, \citenamefont {Mattevi},\ and\ \citenamefont
  {Bouchiat}}]{wakamura2018strong}%
  \BibitemOpen
  \bibfield  {author} {\bibinfo {author} {\bibfnamefont {T.}~\bibnamefont
  {Wakamura}}, \bibinfo {author} {\bibfnamefont {F.}~\bibnamefont {Reale}},
  \bibinfo {author} {\bibfnamefont {P.}~\bibnamefont {Palczynski}}, \bibinfo
  {author} {\bibfnamefont {S.}~\bibnamefont {Gu\'eron}}, \bibinfo {author}
  {\bibfnamefont {C.}~\bibnamefont {Mattevi}}, \ and\ \bibinfo {author}
  {\bibfnamefont {H.}~\bibnamefont {Bouchiat}},\ }\href {\doibase
  10.1103/PhysRevLett.120.106802} {\bibfield  {journal} {\bibinfo  {journal}
  {Phys. Rev. Lett.}\ }\textbf {\bibinfo {volume} {120}},\ \bibinfo {pages}
  {106802} (\bibinfo {year} {2018})}\BibitemShut {NoStop}%
\bibitem [{\citenamefont {Lu}\ \emph {et~al.}(2013)\citenamefont {Lu},
  \citenamefont {Yao}, \citenamefont {Xiao},\ and\ \citenamefont
  {Shen}}]{lu2013intervalley}%
  \BibitemOpen
  \bibfield  {author} {\bibinfo {author} {\bibfnamefont {H.-Z.}\ \bibnamefont
  {Lu}}, \bibinfo {author} {\bibfnamefont {W.}~\bibnamefont {Yao}}, \bibinfo
  {author} {\bibfnamefont {D.}~\bibnamefont {Xiao}}, \ and\ \bibinfo {author}
  {\bibfnamefont {S.-Q.}\ \bibnamefont {Shen}},\ }\href {\doibase
  10.1103/PhysRevLett.110.016806} {\bibfield  {journal} {\bibinfo  {journal}
  {Phys. Rev. Lett.}\ }\textbf {\bibinfo {volume} {110}},\ \bibinfo {pages}
  {016806} (\bibinfo {year} {2013})}\BibitemShut {NoStop}%
\bibitem [{\citenamefont {Imura}\ \emph {et~al.}(2009)\citenamefont {Imura},
  \citenamefont {Kuramoto},\ and\ \citenamefont {Nomura}}]{imura2009weak}%
  \BibitemOpen
  \bibfield  {author} {\bibinfo {author} {\bibfnamefont {K.-I.}\ \bibnamefont
  {Imura}}, \bibinfo {author} {\bibfnamefont {Y.}~\bibnamefont {Kuramoto}}, \
  and\ \bibinfo {author} {\bibfnamefont {K.}~\bibnamefont {Nomura}},\ }\href
  {\doibase 10.1103/PhysRevB.80.085119} {\bibfield  {journal} {\bibinfo
  {journal} {Phys. Rev. B}\ }\textbf {\bibinfo {volume} {80}},\ \bibinfo
  {pages} {085119} (\bibinfo {year} {2009})}\BibitemShut {NoStop}%
\bibitem [{\citenamefont {Shan}\ \emph {et~al.}(2012)\citenamefont {Shan},
  \citenamefont {Lu},\ and\ \citenamefont {Shen}}]{shan2012spin}%
  \BibitemOpen
  \bibfield  {author} {\bibinfo {author} {\bibfnamefont {W.-Y.}\ \bibnamefont
  {Shan}}, \bibinfo {author} {\bibfnamefont {H.-Z.}\ \bibnamefont {Lu}}, \ and\
  \bibinfo {author} {\bibfnamefont {S.-Q.}\ \bibnamefont {Shen}},\ }\href
  {\doibase 10.1103/PhysRevB.86.125303} {\bibfield  {journal} {\bibinfo
  {journal} {Phys. Rev. B}\ }\textbf {\bibinfo {volume} {86}},\ \bibinfo
  {pages} {125303} (\bibinfo {year} {2012})}\BibitemShut {NoStop}%
\bibitem [{\citenamefont {Lu}\ and\ \citenamefont {Shen}(2011)}]{lu2011weak}%
  \BibitemOpen
  \bibfield  {author} {\bibinfo {author} {\bibfnamefont {H.-Z.}\ \bibnamefont
  {Lu}}\ and\ \bibinfo {author} {\bibfnamefont {S.-Q.}\ \bibnamefont {Shen}},\
  }\href {\doibase 10.1103/PhysRevB.84.125138} {\bibfield  {journal} {\bibinfo
  {journal} {Phys. Rev. B}\ }\textbf {\bibinfo {volume} {84}},\ \bibinfo
  {pages} {125138} (\bibinfo {year} {2011})}\BibitemShut {NoStop}%
\bibitem [{\citenamefont {Lu}\ \emph {et~al.}(2011)\citenamefont {Lu},
  \citenamefont {Shi},\ and\ \citenamefont {Shen}}]{lu2011competition}%
  \BibitemOpen
  \bibfield  {author} {\bibinfo {author} {\bibfnamefont {H.-Z.}\ \bibnamefont
  {Lu}}, \bibinfo {author} {\bibfnamefont {J.}~\bibnamefont {Shi}}, \ and\
  \bibinfo {author} {\bibfnamefont {S.-Q.}\ \bibnamefont {Shen}},\ }\href
  {\doibase 10.1103/PhysRevLett.107.076801} {\bibfield  {journal} {\bibinfo
  {journal} {Phys. Rev. Lett.}\ }\textbf {\bibinfo {volume} {107}},\ \bibinfo
  {pages} {076801} (\bibinfo {year} {2011})}\BibitemShut {NoStop}%
\bibitem [{\citenamefont {Cao}\ \emph {et~al.}(2018{\natexlab{a}})\citenamefont
  {Cao}, \citenamefont {Fatemi}, \citenamefont {Demir}, \citenamefont {Fang},
  \citenamefont {Tomarken}, \citenamefont {Luo}, \citenamefont
  {Sanchez-Yamagishi}, \citenamefont {Watanabe}, \citenamefont {Taniguchi},
  \citenamefont {Kaxiras}, \citenamefont {Ashoori},\ and\ \citenamefont
  {Jarillo-Herrero}}]{cao2018correlated}%
  \BibitemOpen
  \bibfield  {author} {\bibinfo {author} {\bibfnamefont {Y.}~\bibnamefont
  {Cao}}, \bibinfo {author} {\bibfnamefont {V.}~\bibnamefont {Fatemi}},
  \bibinfo {author} {\bibfnamefont {A.}~\bibnamefont {Demir}}, \bibinfo
  {author} {\bibfnamefont {S.}~\bibnamefont {Fang}}, \bibinfo {author}
  {\bibfnamefont {S.~L.}\ \bibnamefont {Tomarken}}, \bibinfo {author}
  {\bibfnamefont {J.~Y.}\ \bibnamefont {Luo}}, \bibinfo {author} {\bibfnamefont
  {J.~D.}\ \bibnamefont {Sanchez-Yamagishi}}, \bibinfo {author} {\bibfnamefont
  {K.}~\bibnamefont {Watanabe}}, \bibinfo {author} {\bibfnamefont
  {T.}~\bibnamefont {Taniguchi}}, \bibinfo {author} {\bibfnamefont
  {E.}~\bibnamefont {Kaxiras}}, \bibinfo {author} {\bibfnamefont {R.~C.}\
  \bibnamefont {Ashoori}}, \ and\ \bibinfo {author} {\bibfnamefont
  {P.}~\bibnamefont {Jarillo-Herrero}},\ }\href {\doibase 10.1038/nature26154}
  {\bibfield  {journal} {\bibinfo  {journal} {Nature}\ }\textbf {\bibinfo
  {volume} {556}},\ \bibinfo {pages} {80} (\bibinfo {year}
  {2018}{\natexlab{a}})}\BibitemShut {NoStop}%
\bibitem [{\citenamefont {Cao}\ \emph {et~al.}(2018{\natexlab{b}})\citenamefont
  {Cao}, \citenamefont {Fatemi}, \citenamefont {Fang}, \citenamefont
  {Watanabe}, \citenamefont {Taniguchi}, \citenamefont {Kaxiras},\ and\
  \citenamefont {Jarillo-Herrero}}]{cao2018unconventional}%
  \BibitemOpen
  \bibfield  {author} {\bibinfo {author} {\bibfnamefont {Y.}~\bibnamefont
  {Cao}}, \bibinfo {author} {\bibfnamefont {V.}~\bibnamefont {Fatemi}},
  \bibinfo {author} {\bibfnamefont {S.}~\bibnamefont {Fang}}, \bibinfo {author}
  {\bibfnamefont {K.}~\bibnamefont {Watanabe}}, \bibinfo {author}
  {\bibfnamefont {T.}~\bibnamefont {Taniguchi}}, \bibinfo {author}
  {\bibfnamefont {E.}~\bibnamefont {Kaxiras}}, \ and\ \bibinfo {author}
  {\bibfnamefont {P.}~\bibnamefont {Jarillo-Herrero}},\ }\href {\doibase
  10.1038/nature26160} {\bibfield  {journal} {\bibinfo  {journal} {Nature}\
  }\textbf {\bibinfo {volume} {556}},\ \bibinfo {pages} {43} (\bibinfo {year}
  {2018}{\natexlab{b}})}\BibitemShut {NoStop}%
\bibitem [{\citenamefont {Lu}\ \emph {et~al.}(2019)\citenamefont {Lu},
  \citenamefont {Stepanov}, \citenamefont {Yang}, \citenamefont {Xie},
  \citenamefont {Aamir}, \citenamefont {Das}, \citenamefont {Urgell},
  \citenamefont {Watanabe}, \citenamefont {Taniguchi}, \citenamefont {Zhang},
  \citenamefont {Bachtold}, \citenamefont {MacDonald},\ and\ \citenamefont
  {Efetov}}]{lu2019superconductors}%
  \BibitemOpen
  \bibfield  {author} {\bibinfo {author} {\bibfnamefont {X.}~\bibnamefont
  {Lu}}, \bibinfo {author} {\bibfnamefont {P.}~\bibnamefont {Stepanov}},
  \bibinfo {author} {\bibfnamefont {W.}~\bibnamefont {Yang}}, \bibinfo {author}
  {\bibfnamefont {M.}~\bibnamefont {Xie}}, \bibinfo {author} {\bibfnamefont
  {M.~A.}\ \bibnamefont {Aamir}}, \bibinfo {author} {\bibfnamefont
  {I.}~\bibnamefont {Das}}, \bibinfo {author} {\bibfnamefont {C.}~\bibnamefont
  {Urgell}}, \bibinfo {author} {\bibfnamefont {K.}~\bibnamefont {Watanabe}},
  \bibinfo {author} {\bibfnamefont {T.}~\bibnamefont {Taniguchi}}, \bibinfo
  {author} {\bibfnamefont {G.}~\bibnamefont {Zhang}}, \bibinfo {author}
  {\bibfnamefont {A.}~\bibnamefont {Bachtold}}, \bibinfo {author}
  {\bibfnamefont {A.~H.}\ \bibnamefont {MacDonald}}, \ and\ \bibinfo {author}
  {\bibfnamefont {D.~K.}\ \bibnamefont {Efetov}},\ }\href {\doibase
  10.1038/s41586-019-1695-0} {\bibfield  {journal} {\bibinfo  {journal}
  {Nature}\ }\textbf {\bibinfo {volume} {574}},\ \bibinfo {pages} {653}
  (\bibinfo {year} {2019})}\BibitemShut {NoStop}%
\bibitem [{\citenamefont {Yankowitz}\ \emph {et~al.}(2019)\citenamefont
  {Yankowitz}, \citenamefont {Chen}, \citenamefont {Polshyn}, \citenamefont
  {Zhang}, \citenamefont {Watanabe}, \citenamefont {Taniguchi}, \citenamefont
  {Graf}, \citenamefont {Young},\ and\ \citenamefont
  {Dean}}]{yankowitz2019tuning}%
  \BibitemOpen
  \bibfield  {author} {\bibinfo {author} {\bibfnamefont {M.}~\bibnamefont
  {Yankowitz}}, \bibinfo {author} {\bibfnamefont {S.}~\bibnamefont {Chen}},
  \bibinfo {author} {\bibfnamefont {H.}~\bibnamefont {Polshyn}}, \bibinfo
  {author} {\bibfnamefont {Y.}~\bibnamefont {Zhang}}, \bibinfo {author}
  {\bibfnamefont {K.}~\bibnamefont {Watanabe}}, \bibinfo {author}
  {\bibfnamefont {T.}~\bibnamefont {Taniguchi}}, \bibinfo {author}
  {\bibfnamefont {D.}~\bibnamefont {Graf}}, \bibinfo {author} {\bibfnamefont
  {A.~F.}\ \bibnamefont {Young}}, \ and\ \bibinfo {author} {\bibfnamefont
  {C.~R.}\ \bibnamefont {Dean}},\ }\href {\doibase 10.1126/science.aav1910}
  {\bibfield  {journal} {\bibinfo  {journal} {Science}\ }\textbf {\bibinfo
  {volume} {363}},\ \bibinfo {pages} {1059} (\bibinfo {year}
  {2019})}\BibitemShut {NoStop}%
\bibitem [{\citenamefont {Bistritzer}\ and\ \citenamefont
  {MacDonald}(2011)}]{bistritzer2011moire}%
  \BibitemOpen
  \bibfield  {author} {\bibinfo {author} {\bibfnamefont {R.}~\bibnamefont
  {Bistritzer}}\ and\ \bibinfo {author} {\bibfnamefont {A.~H.}\ \bibnamefont
  {MacDonald}},\ }\href {\doibase 10.1073/pnas.1108174108} {\bibfield
  {journal} {\bibinfo  {journal} {Proceedings of the National Academy of
  Sciences}\ }\textbf {\bibinfo {volume} {108}},\ \bibinfo {pages} {12233}
  (\bibinfo {year} {2011})}\BibitemShut {NoStop}%
\bibitem [{\citenamefont {Lopes~dos Santos}\ \emph {et~al.}(2007)\citenamefont
  {Lopes~dos Santos}, \citenamefont {Peres},\ and\ \citenamefont
  {Castro~Neto}}]{dos2007graphene}%
  \BibitemOpen
  \bibfield  {author} {\bibinfo {author} {\bibfnamefont {J.~M.~B.}\
  \bibnamefont {Lopes~dos Santos}}, \bibinfo {author} {\bibfnamefont
  {N.~M.~R.}\ \bibnamefont {Peres}}, \ and\ \bibinfo {author} {\bibfnamefont
  {A.~H.}\ \bibnamefont {Castro~Neto}},\ }\href {\doibase
  10.1103/PhysRevLett.99.256802} {\bibfield  {journal} {\bibinfo  {journal}
  {Phys. Rev. Lett.}\ }\textbf {\bibinfo {volume} {99}},\ \bibinfo {pages}
  {256802} (\bibinfo {year} {2007})}\BibitemShut {NoStop}%
\bibitem [{\citenamefont {Moon}\ and\ \citenamefont
  {Koshino}(2013)}]{moon2013optical}%
  \BibitemOpen
  \bibfield  {author} {\bibinfo {author} {\bibfnamefont {P.}~\bibnamefont
  {Moon}}\ and\ \bibinfo {author} {\bibfnamefont {M.}~\bibnamefont {Koshino}},\
  }\href {\doibase 10.1103/PhysRevB.87.205404} {\bibfield  {journal} {\bibinfo
  {journal} {Phys. Rev. B}\ }\textbf {\bibinfo {volume} {87}},\ \bibinfo
  {pages} {205404} (\bibinfo {year} {2013})}\BibitemShut {NoStop}%
\bibitem [{\citenamefont {Koshino}\ \emph {et~al.}(2018)\citenamefont
  {Koshino}, \citenamefont {Yuan}, \citenamefont {Koretsune}, \citenamefont
  {Ochi}, \citenamefont {Kuroki},\ and\ \citenamefont
  {Fu}}]{koshino2018maximally}%
  \BibitemOpen
  \bibfield  {author} {\bibinfo {author} {\bibfnamefont {M.}~\bibnamefont
  {Koshino}}, \bibinfo {author} {\bibfnamefont {N.~F.~Q.}\ \bibnamefont
  {Yuan}}, \bibinfo {author} {\bibfnamefont {T.}~\bibnamefont {Koretsune}},
  \bibinfo {author} {\bibfnamefont {M.}~\bibnamefont {Ochi}}, \bibinfo {author}
  {\bibfnamefont {K.}~\bibnamefont {Kuroki}}, \ and\ \bibinfo {author}
  {\bibfnamefont {L.}~\bibnamefont {Fu}},\ }\href {\doibase
  10.1103/PhysRevX.8.031087} {\bibfield  {journal} {\bibinfo  {journal} {Phys.
  Rev. X}\ }\textbf {\bibinfo {volume} {8}},\ \bibinfo {pages} {031087}
  (\bibinfo {year} {2018})}\BibitemShut {NoStop}%
\bibitem [{\citenamefont {Meng}\ \emph {et~al.}(2012)\citenamefont {Meng},
  \citenamefont {Chu}, \citenamefont {Zhang}, \citenamefont {Yang},
  \citenamefont {Dou}, \citenamefont {Nie},\ and\ \citenamefont
  {He}}]{meng2012enhanced}%
  \BibitemOpen
  \bibfield  {author} {\bibinfo {author} {\bibfnamefont {L.}~\bibnamefont
  {Meng}}, \bibinfo {author} {\bibfnamefont {Z.-D.}\ \bibnamefont {Chu}},
  \bibinfo {author} {\bibfnamefont {Y.}~\bibnamefont {Zhang}}, \bibinfo
  {author} {\bibfnamefont {J.-Y.}\ \bibnamefont {Yang}}, \bibinfo {author}
  {\bibfnamefont {R.-F.}\ \bibnamefont {Dou}}, \bibinfo {author} {\bibfnamefont
  {J.-C.}\ \bibnamefont {Nie}}, \ and\ \bibinfo {author} {\bibfnamefont
  {L.}~\bibnamefont {He}},\ }\href {\doibase 10.1103/PhysRevB.85.235453}
  {\bibfield  {journal} {\bibinfo  {journal} {Phys. Rev. B}\ }\textbf {\bibinfo
  {volume} {85}},\ \bibinfo {pages} {235453} (\bibinfo {year}
  {2012})}\BibitemShut {NoStop}%
\bibitem [{\citenamefont {Kashiwagi}\ \emph {et~al.}(2022)\citenamefont
  {Kashiwagi}, \citenamefont {Taen}, \citenamefont {Uchida}, \citenamefont
  {Watanabe}, \citenamefont {Taniguchi},\ and\ \citenamefont
  {Osada}}]{kashiwagi2022weak}%
  \BibitemOpen
  \bibfield  {author} {\bibinfo {author} {\bibfnamefont {M.}~\bibnamefont
  {Kashiwagi}}, \bibinfo {author} {\bibfnamefont {T.}~\bibnamefont {Taen}},
  \bibinfo {author} {\bibfnamefont {K.}~\bibnamefont {Uchida}}, \bibinfo
  {author} {\bibfnamefont {K.}~\bibnamefont {Watanabe}}, \bibinfo {author}
  {\bibfnamefont {T.}~\bibnamefont {Taniguchi}}, \ and\ \bibinfo {author}
  {\bibfnamefont {T.}~\bibnamefont {Osada}},\ }\href {\doibase
  10.35848/1347-4065/ac934a} {\bibfield  {journal} {\bibinfo  {journal} {Jap.
  J. Appl. Phys}\ }\textbf {\bibinfo {volume} {61}},\ \bibinfo {pages} {100907}
  (\bibinfo {year} {2022})}\BibitemShut {NoStop}%
\bibitem [{\citenamefont {Phinney}\ \emph {et~al.}(2021)\citenamefont
  {Phinney}, \citenamefont {Bandurin}, \citenamefont {Collignon}, \citenamefont
  {Dmitriev}, \citenamefont {Taniguchi}, \citenamefont {Watanabe},\ and\
  \citenamefont {Jarillo-Herrero}}]{phinney2021strong}%
  \BibitemOpen
  \bibfield  {author} {\bibinfo {author} {\bibfnamefont {I.~Y.}\ \bibnamefont
  {Phinney}}, \bibinfo {author} {\bibfnamefont {D.~A.}\ \bibnamefont
  {Bandurin}}, \bibinfo {author} {\bibfnamefont {C.}~\bibnamefont {Collignon}},
  \bibinfo {author} {\bibfnamefont {I.~A.}\ \bibnamefont {Dmitriev}}, \bibinfo
  {author} {\bibfnamefont {T.}~\bibnamefont {Taniguchi}}, \bibinfo {author}
  {\bibfnamefont {K.}~\bibnamefont {Watanabe}}, \ and\ \bibinfo {author}
  {\bibfnamefont {P.}~\bibnamefont {Jarillo-Herrero}},\ }\href {\doibase
  10.1103/PhysRevLett.127.056802} {\bibfield  {journal} {\bibinfo  {journal}
  {Phys. Rev. Lett.}\ }\textbf {\bibinfo {volume} {127}},\ \bibinfo {pages}
  {056802} (\bibinfo {year} {2021})}\BibitemShut {NoStop}%
\bibitem [{\citenamefont {Meyer}\ \emph {et~al.}(2007)\citenamefont {Meyer},
  \citenamefont {Geim}, \citenamefont {Katsnelson}, \citenamefont {Novoselov},
  \citenamefont {Booth},\ and\ \citenamefont {Roth}}]{meyer2007structure}%
  \BibitemOpen
  \bibfield  {author} {\bibinfo {author} {\bibfnamefont {J.~C.}\ \bibnamefont
  {Meyer}}, \bibinfo {author} {\bibfnamefont {A.~K.}\ \bibnamefont {Geim}},
  \bibinfo {author} {\bibfnamefont {M.~I.}\ \bibnamefont {Katsnelson}},
  \bibinfo {author} {\bibfnamefont {K.~S.}\ \bibnamefont {Novoselov}}, \bibinfo
  {author} {\bibfnamefont {T.~J.}\ \bibnamefont {Booth}}, \ and\ \bibinfo
  {author} {\bibfnamefont {S.}~\bibnamefont {Roth}},\ }\href {\doibase
  10.1038/nature05545} {\bibfield  {journal} {\bibinfo  {journal} {Nature}\
  }\textbf {\bibinfo {volume} {446}},\ \bibinfo {pages} {60} (\bibinfo {year}
  {2007})}\BibitemShut {NoStop}%
\bibitem [{\citenamefont {Skomski}\ \emph {et~al.}(2014)\citenamefont
  {Skomski}, \citenamefont {Dowben}, \citenamefont {Driver},\ and\
  \citenamefont {Kelber}}]{skomski2014sublattice}%
  \BibitemOpen
  \bibfield  {author} {\bibinfo {author} {\bibfnamefont {R.}~\bibnamefont
  {Skomski}}, \bibinfo {author} {\bibfnamefont {P.~A.}\ \bibnamefont {Dowben}},
  \bibinfo {author} {\bibfnamefont {M.~S.}\ \bibnamefont {Driver}}, \ and\
  \bibinfo {author} {\bibfnamefont {J.~A.}\ \bibnamefont {Kelber}},\ }\href
  {\doibase 10.1039/c4mh00124a} {\bibfield  {journal} {\bibinfo  {journal}
  {Mater. Horizons}\ }\textbf {\bibinfo {volume} {1}},\ \bibinfo {pages} {563}
  (\bibinfo {year} {2014})}\BibitemShut {NoStop}%
\bibitem [{\citenamefont {Tomi\ifmmode~\acute{c}\else \'{c}\fi{}}\ \emph
  {et~al.}(2022)\citenamefont {Tomi\ifmmode~\acute{c}\else \'{c}\fi{}},
  \citenamefont {Rickhaus}, \citenamefont {Garcia-Ruiz}, \citenamefont {Zheng},
  \citenamefont {Portol\'es}, \citenamefont {Fal'ko}, \citenamefont {Watanabe},
  \citenamefont {Taniguchi}, \citenamefont {Ensslin}, \citenamefont {Ihn},\
  and\ \citenamefont {de~Vries}}]{tomic2022scattering}%
  \BibitemOpen
  \bibfield  {author} {\bibinfo {author} {\bibfnamefont {P.}~\bibnamefont
  {Tomi\ifmmode~\acute{c}\else \'{c}\fi{}}}, \bibinfo {author} {\bibfnamefont
  {P.}~\bibnamefont {Rickhaus}}, \bibinfo {author} {\bibfnamefont
  {A.}~\bibnamefont {Garcia-Ruiz}}, \bibinfo {author} {\bibfnamefont
  {G.}~\bibnamefont {Zheng}}, \bibinfo {author} {\bibfnamefont
  {E.}~\bibnamefont {Portol\'es}}, \bibinfo {author} {\bibfnamefont
  {V.}~\bibnamefont {Fal'ko}}, \bibinfo {author} {\bibfnamefont
  {K.}~\bibnamefont {Watanabe}}, \bibinfo {author} {\bibfnamefont
  {T.}~\bibnamefont {Taniguchi}}, \bibinfo {author} {\bibfnamefont
  {K.}~\bibnamefont {Ensslin}}, \bibinfo {author} {\bibfnamefont
  {T.}~\bibnamefont {Ihn}}, \ and\ \bibinfo {author} {\bibfnamefont {F.~K.}\
  \bibnamefont {de~Vries}},\ }\href {\doibase 10.1103/PhysRevLett.128.057702}
  {\bibfield  {journal} {\bibinfo  {journal} {Phys. Rev. Lett.}\ }\textbf
  {\bibinfo {volume} {128}},\ \bibinfo {pages} {057702} (\bibinfo {year}
  {2022})}\BibitemShut {NoStop}%
\bibitem [{\citenamefont {Akkermans}\ and\ \citenamefont
  {Montambaux}(2007)}]{akkermans2007mesoscopic}%
  \BibitemOpen
  \bibfield  {author} {\bibinfo {author} {\bibfnamefont {E.}~\bibnamefont
  {Akkermans}}\ and\ \bibinfo {author} {\bibfnamefont {G.}~\bibnamefont
  {Montambaux}},\ }\href@noop {} {\emph {\bibinfo {title} {Mesoscopic physics
  of electrons and photons}}}\ (\bibinfo  {publisher} {Cambridge university
  press},\ \bibinfo {year} {2007})\BibitemShut {NoStop}%
\bibitem [{\citenamefont {Datta}(1997)}]{datta1997electronic}%
  \BibitemOpen
  \bibfield  {author} {\bibinfo {author} {\bibfnamefont {S.}~\bibnamefont
  {Datta}},\ }\href@noop {} {\emph {\bibinfo {title} {Electronic transport in
  mesoscopic systems}}}\ (\bibinfo  {publisher} {Cambridge university press},\
  \bibinfo {year} {1997})\BibitemShut {NoStop}%
\end{thebibliography}%
\end{document}